\documentclass[longauth]{aa}
\usepackage{euclid}

\usepackage{amsfonts}
\usepackage{graphicx}
\usepackage{txfonts}
\usepackage[pdfencoding=auto,psdextra]{hyperref}
\hypersetup{
    colorlinks=true,
    linkcolor=blue,
    filecolor=magenta,      
    urlcolor=blue,
    citecolor=blue
}
\usepackage{color, colortbl}
\usepackage{placeins}
\usepackage{tabularx}
\usepackage{comment}
\usepackage{subcaption}
\usepackage{multirow}
\usepackage{lastpage}
\usepackage[nameinlink,capitalise]{cleveref}
\usepackage{acro}
\usepackage{lscape}

\crefname{section}{Sect.}{Sects.}
\Crefname{section}{Section}{Sections}
\crefname{figure}{Fig.}{Figs.}
\Crefname{figure}{Figure}{Figures}
\crefname{equation}{Eq.}{Eqs.}
\Crefname{equation}{Equation}{Equations}
\crefname{table}{Table}{Tables}
\crefname{appendix}{Appendix}{Appendices}

\newcommand{\BGlow}{\ensuremath{926}}
\newcommand{\BGhigh}{\ensuremath{1366}}
\newcommand{\RGlow}{\ensuremath{1206}}
\newcommand{\RGhigh}{\ensuremath{1892}}

\ifdefined\BGE
    \renewcommand{\BGE}{\ensuremath{BG_\sfont{E}}}
\else
    \newcommand{\BGE}{\ensuremath{BG_\sfont{E}}}
\fi

\ifdefined\RGE
    \renewcommand{\RGE}{\ensuremath{RG_\sfont{E}}}
\else
    \newcommand{\RGE}{\ensuremath{RG_\sfont{E}}}
\fi

\newcommand{\ShortTitle}{The NISP spectroscopy channel, on ground performance and calibration}

\DeclareAcronym{CoLA}{short=CoLA,long=corrector lens}
\DeclareAcronym{CaLA}{short=CaLA,long=camera lens assembly}
\DeclareAcronym{CPPM}{short=CPPM,long={`Centre de physique des particules de Marseille'}}
\DeclareAcronym{CSL}{short=CSL,long={`Centre Spatial de Liège'}}
\DeclareAcronym{ESA}{short=ESA,long=European Space Agency}
\DeclareAcronym{EE}{short=EE,long=encircled energy}
\DeclareAcronym{FoM}{short=FoM,long=Figure of Merit}
\DeclareAcronym{FoV}{short=FoV,long=field-of-view}
\DeclareAcronym{FWHM}{short=FWHM,long=Full-Width-Half-Maximum}
\DeclareAcronym{IPC}{short=IPC,long=Intra Pixel Capacitance}
\DeclareAcronym{IP2I}{short=IP2I,long={`Institut des deux infinis'}}
\DeclareAcronym{LAM}{short=LAM,long={`Laboratoire d'Astrophysique de Marseille'}}
\DeclareAcronym{MHP}{short=MHP,long=methane heat pipes}
\DeclareAcronym{NI-ICU}{short=NI-ICU,long=Instrument Control Unit}
\DeclareAcronym{NIR}{short=NIR,long=near-infrared}
\DeclareAcronym{NISP}{short=NISP,long=Near-Infrared Spectrometer and Photometer}
\DeclareAcronym{NI-CU}{short=NI-CU,long=calibration unit}
\DeclareAcronym{NI-DCU}{short=NI-DCU,long=Data Control Unit}
\DeclareAcronym{NI-DPU}{short=NI-DPU,long=Data Processing Unit}
\DeclareAcronym{NI-DS}{short=NI-DS,long=detector system}
\DeclareAcronym{NI-FWA}{short=NI-FWA,long=filter wheel}
\DeclareAcronym{NI-GWA}{short=NI-GWA,long=grism wheel}
\DeclareAcronym{NI-OA}{short=NI-OA,long=NISP Optical Assembly}
\DeclareAcronym{NI-SA-HP}{short=NI-SA-HP,long=NISP Structure Assembly HexaPods}
\DeclareAcronym{NI-SA-ST}{short=NI-SA-ST,long=NISP Structure Assembly STructure}
\DeclareAcronym{PLM}{short=PLM,long=Payload Module}
\DeclareAcronym{PSF}{short=PSF,long=point spread function}
\DeclareAcronym{SED}{short=SED,long=spectral energy distribution}
\DeclareAcronym{SiC}{short=SiC,long=Silicon Carbide}

\begin{document}

\title{\Euclid preparation}
\subtitle{The NISP spectroscopy channel, on ground performance and calibration}    

\newcommand{\orcid}[1]{\protect\href{https://orcid.org/#1}{\protect\includegraphics[width=6pt]{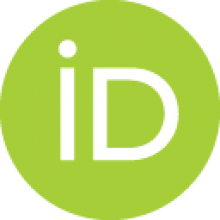}}}

\author{Euclid Collaboration: W.~Gillard\orcid{0000-0003-4744-9748}\thanks{\email{gillard@cppm.in2P3.fr}}\inst{\ref{aff1}}
\and T.~Maciaszek\inst{\ref{aff2}}
\and E.~Prieto\inst{\ref{aff3}}
\and F.~Grupp\inst{\ref{aff4},\ref{aff5}}
\and A.~Costille\inst{\ref{aff3}}
\and K.~Jahnke\orcid{0000-0003-3804-2137}\inst{\ref{aff6}}
\and J.~Clemens\inst{\ref{aff1}}
\and S.~Dusini\orcid{0000-0002-1128-0664}\inst{\ref{aff7}}
\and M.~Carle\inst{\ref{aff3}}
\and C.~Sirignano\orcid{0000-0002-0995-7146}\inst{\ref{aff8},\ref{aff7}}
\and E.~Medinaceli\orcid{0000-0002-4040-7783}\inst{\ref{aff9}}
\and S.~Ligori\orcid{0000-0003-4172-4606}\inst{\ref{aff10}}
\and E.~Franceschi\orcid{0000-0002-0585-6591}\inst{\ref{aff9}}
\and M.~Trifoglio\orcid{0000-0002-2505-3630}\inst{\ref{aff9}}
\and W.~Bon\inst{\ref{aff3}}
\and R.~Barbier\inst{\ref{aff11}}
\and S.~Ferriol\inst{\ref{aff11}}
\and A.~Secroun\orcid{0000-0003-0505-3710}\inst{\ref{aff1}}
\and N.~Auricchio\orcid{0000-0003-4444-8651}\inst{\ref{aff9}}
\and P.~Battaglia\orcid{0000-0002-7337-5909}\inst{\ref{aff9}}
\and C.~Bonoli\inst{\ref{aff12}}
\and L.~Corcione\orcid{0000-0002-6497-5881}\inst{\ref{aff10}}
\and F.~Hormuth\inst{\ref{aff13}}
\and D.~Le~Mignant\orcid{0000-0002-5339-5515}\inst{\ref{aff3}}
\and G.~Morgante\inst{\ref{aff9}}
\and C.~Padilla\orcid{0000-0001-7951-0166}\inst{\ref{aff14}}
\and R.~Toledo-Moreo\orcid{0000-0002-2997-4859}\inst{\ref{aff15}}
\and L.~Valenziano\orcid{0000-0002-1170-0104}\inst{\ref{aff9},\ref{aff16}}
\and R.~Bender\orcid{0000-0001-7179-0626}\inst{\ref{aff4},\ref{aff5}}
\and F.~J.~Castander\orcid{0000-0001-7316-4573}\inst{\ref{aff17},\ref{aff18}}
\and P.~B.~Lilje\orcid{0000-0003-4324-7794}\inst{\ref{aff19}}
\and A.~Balestra\orcid{0000-0002-6967-261X}\inst{\ref{aff12}}
\and J.-J.~C.~Barriere\inst{\ref{aff20}}
\and M.~Berthe\inst{\ref{aff21}}
\and C.~Boderndorf\inst{\ref{aff4}}
\and A.~Bonnefoi\inst{\ref{aff3}}
\and V.~Capobianco\orcid{0000-0002-3309-7692}\inst{\ref{aff10}}
\and R.~Casas\orcid{0000-0002-8165-5601}\inst{\ref{aff18},\ref{aff17}}
\and H.~Cho\inst{\ref{aff22}}
\and F.~Ducret\inst{\ref{aff3}}
\and J.-L.~Gimenez\inst{\ref{aff3}}
\and W.~Holmes\inst{\ref{aff22}}
\and A.~Hornstrup\orcid{0000-0002-3363-0936}\inst{\ref{aff23},\ref{aff24}}
\and M.~Jhabvala\inst{\ref{aff25}}
\and E.~Jullo\orcid{0000-0002-9253-053X}\inst{\ref{aff3}}
\and R.~Kohley\inst{\ref{aff26}}
\and B.~Kubik\orcid{0009-0006-5823-4880}\inst{\ref{aff11}}
\and R.~Laureijs\inst{\ref{aff27},\ref{aff28}}
\and I.~Lloro\orcid{0000-0001-5966-1434}\inst{\ref{aff29}}
\and C.~Macabiau\inst{\ref{aff11}}
\and Y.~Mellier\inst{\ref{aff30},\ref{aff31}}
\and G.~Polenta\orcid{0000-0003-4067-9196}\inst{\ref{aff32}}
\and G.~D.~Racca\inst{\ref{aff27},\ref{aff33}}
\and A.~Renzi\orcid{0000-0001-9856-1970}\inst{\ref{aff8},\ref{aff7}}
\and M.~Schirmer\orcid{0000-0003-2568-9994}\inst{\ref{aff6}}
\and G.~Seidel\orcid{0000-0003-2907-353X}\inst{\ref{aff6}}
\and M.~Seiffert\orcid{0000-0002-7536-9393}\inst{\ref{aff22}}
\and G.~Sirri\orcid{0000-0003-2626-2853}\inst{\ref{aff34}}
\and G.~Smadja\inst{\ref{aff11}}
\and L.~Stanco\orcid{0000-0002-9706-5104}\inst{\ref{aff7}}
\and S.~Wachter\inst{\ref{aff35}}
\and H.~Aussel\orcid{0000-0002-1371-5705}\inst{\ref{aff21}}
\and T.~Auphan\orcid{0009-0008-9988-3646}\inst{\ref{aff1}}
\and B.~R.~Granett\orcid{0000-0003-2694-9284}\inst{\ref{aff36}}
\and R.~Chary\orcid{0000-0001-7583-0621}\inst{\ref{aff37},\ref{aff38}}
\and Y.~Copin\orcid{0000-0002-5317-7518}\inst{\ref{aff11}}
\and P.~Hudelot\inst{\ref{aff31}}
\and V.~Le~Brun\orcid{0000-0002-5027-1939}\inst{\ref{aff3}}
\and F.~Torradeflot\orcid{0000-0003-1160-1517}\inst{\ref{aff39},\ref{aff40}}
\and P.~N.~Appleton\orcid{0000-0002-7607-8766}\inst{\ref{aff41}}
\and P.~Casenove\orcid{0009-0006-6736-1670}\inst{\ref{aff2}}
\and P.-Y.~Chabaud\inst{\ref{aff3}}
\and M.~Frailis\orcid{0000-0002-7400-2135}\inst{\ref{aff42}}
\and M.~Fumana\orcid{0000-0001-6787-5950}\inst{\ref{aff43}}
\and L.~Guzzo\orcid{0000-0001-8264-5192}\inst{\ref{aff44},\ref{aff36},\ref{aff45}}
\and G.~Mainetti\orcid{0000-0003-2384-2377}\inst{\ref{aff46}}
\and D.~Maino\inst{\ref{aff44},\ref{aff43},\ref{aff45}}
\and M.~Moresco\orcid{0000-0002-7616-7136}\inst{\ref{aff47},\ref{aff9}}
\and W.~J.~Percival\orcid{0000-0002-0644-5727}\inst{\ref{aff48},\ref{aff49},\ref{aff50}}
\and R.~Scaramella\orcid{0000-0003-2229-193X}\inst{\ref{aff51},\ref{aff52}}
\and M.~Scodeggio\inst{\ref{aff43}}
\and N.~R.~Stickley\orcid{0000-0003-0987-5738}\inst{\ref{aff53}}
\and D.~Vibert\orcid{0009-0008-0607-631X}\inst{\ref{aff3}}
\and Y.~Wang\orcid{0000-0002-4749-2984}\inst{\ref{aff37}}
\and J.~Zoubian\inst{\ref{aff1}}
\and N.~Aghanim\orcid{0000-0002-6688-8992}\inst{\ref{aff54}}
\and B.~Altieri\orcid{0000-0003-3936-0284}\inst{\ref{aff26}}
\and A.~Amara\inst{\ref{aff55}}
\and S.~Andreon\orcid{0000-0002-2041-8784}\inst{\ref{aff36}}
\and C.~Baccigalupi\orcid{0000-0002-8211-1630}\inst{\ref{aff56},\ref{aff42},\ref{aff57},\ref{aff58}}
\and M.~Baldi\orcid{0000-0003-4145-1943}\inst{\ref{aff59},\ref{aff9},\ref{aff34}}
\and S.~Bardelli\orcid{0000-0002-8900-0298}\inst{\ref{aff9}}
\and A.~Biviano\orcid{0000-0002-0857-0732}\inst{\ref{aff42},\ref{aff56}}
\and A.~Bonchi\orcid{0000-0002-2667-5482}\inst{\ref{aff32}}
\and E.~Branchini\orcid{0000-0002-0808-6908}\inst{\ref{aff60},\ref{aff61},\ref{aff36}}
\and M.~Brescia\orcid{0000-0001-9506-5680}\inst{\ref{aff62},\ref{aff63}}
\and J.~Brinchmann\orcid{0000-0003-4359-8797}\inst{\ref{aff64},\ref{aff65}}
\and S.~Camera\orcid{0000-0003-3399-3574}\inst{\ref{aff66},\ref{aff67},\ref{aff10}}
\and G.~Ca\~nas-Herrera\orcid{0000-0003-2796-2149}\inst{\ref{aff27},\ref{aff68},\ref{aff33}}
\and C.~Carbone\orcid{0000-0003-0125-3563}\inst{\ref{aff43}}
\and J.~Carretero\orcid{0000-0002-3130-0204}\inst{\ref{aff40},\ref{aff39}}
\and S.~Casas\orcid{0000-0002-4751-5138}\inst{\ref{aff69}}
\and M.~Castellano\orcid{0000-0001-9875-8263}\inst{\ref{aff51}}
\and G.~Castignani\orcid{0000-0001-6831-0687}\inst{\ref{aff9}}
\and S.~Cavuoti\orcid{0000-0002-3787-4196}\inst{\ref{aff63},\ref{aff70}}
\and K.~C.~Chambers\orcid{0000-0001-6965-7789}\inst{\ref{aff71}}
\and A.~Cimatti\inst{\ref{aff72}}
\and C.~Colodro-Conde\inst{\ref{aff73}}
\and G.~Congedo\orcid{0000-0003-2508-0046}\inst{\ref{aff74}}
\and C.~J.~Conselice\orcid{0000-0003-1949-7638}\inst{\ref{aff75}}
\and L.~Conversi\orcid{0000-0002-6710-8476}\inst{\ref{aff76},\ref{aff26}}
\and F.~Courbin\orcid{0000-0003-0758-6510}\inst{\ref{aff77},\ref{aff78}}
\and H.~M.~Courtois\orcid{0000-0003-0509-1776}\inst{\ref{aff79}}
\and J.-G.~Cuby\orcid{0000-0002-8767-1442}\inst{\ref{aff80},\ref{aff3}}
\and A.~Da~Silva\orcid{0000-0002-6385-1609}\inst{\ref{aff81},\ref{aff82}}
\and R.~da~Silva\orcid{0000-0003-4788-677X}\inst{\ref{aff51},\ref{aff32}}
\and H.~Degaudenzi\orcid{0000-0002-5887-6799}\inst{\ref{aff83}}
\and G.~De~Lucia\orcid{0000-0002-6220-9104}\inst{\ref{aff42}}
\and A.~M.~Di~Giorgio\orcid{0000-0002-4767-2360}\inst{\ref{aff84}}
\and H.~Dole\orcid{0000-0002-9767-3839}\inst{\ref{aff54}}
\and M.~Douspis\orcid{0000-0003-4203-3954}\inst{\ref{aff54}}
\and F.~Dubath\orcid{0000-0002-6533-2810}\inst{\ref{aff83}}
\and X.~Dupac\inst{\ref{aff26}}
\and S.~Escoffier\orcid{0000-0002-2847-7498}\inst{\ref{aff1}}
\and M.~Fabricius\orcid{0000-0002-7025-6058}\inst{\ref{aff4},\ref{aff5}}
\and M.~Farina\orcid{0000-0002-3089-7846}\inst{\ref{aff84}}
\and R.~Farinelli\inst{\ref{aff9}}
\and P.~Fosalba\orcid{0000-0002-1510-5214}\inst{\ref{aff18},\ref{aff17}}
\and S.~Fotopoulou\orcid{0000-0002-9686-254X}\inst{\ref{aff85}}
\and N.~Fourmanoit\orcid{0009-0005-6816-6925}\inst{\ref{aff1}}
\and P.~Franzetti\inst{\ref{aff43}}
\and S.~Galeotta\orcid{0000-0002-3748-5115}\inst{\ref{aff42}}
\and K.~George\orcid{0000-0002-1734-8455}\inst{\ref{aff5}}
\and B.~Gillis\orcid{0000-0002-4478-1270}\inst{\ref{aff74}}
\and C.~Giocoli\orcid{0000-0002-9590-7961}\inst{\ref{aff9},\ref{aff34}}
\and P.~G\'omez-Alvarez\orcid{0000-0002-8594-5358}\inst{\ref{aff86},\ref{aff26}}
\and J.~Gracia-Carpio\inst{\ref{aff4}}
\and A.~Grazian\orcid{0000-0002-5688-0663}\inst{\ref{aff12}}
\and S.~V.~H.~Haugan\orcid{0000-0001-9648-7260}\inst{\ref{aff19}}
\and J.~Hoar\inst{\ref{aff26}}
\and H.~Hoekstra\orcid{0000-0002-0641-3231}\inst{\ref{aff33}}
\and I.~M.~Hook\orcid{0000-0002-2960-978X}\inst{\ref{aff87}}
\and E.~Keih\"anen\orcid{0000-0003-1804-7715}\inst{\ref{aff88}}
\and S.~Kermiche\orcid{0000-0002-0302-5735}\inst{\ref{aff1}}
\and A.~Kiessling\orcid{0000-0002-2590-1273}\inst{\ref{aff22}}
\and M.~K\"ummel\orcid{0000-0003-2791-2117}\inst{\ref{aff5}}
\and M.~Kunz\orcid{0000-0002-3052-7394}\inst{\ref{aff89}}
\and H.~Kurki-Suonio\orcid{0000-0002-4618-3063}\inst{\ref{aff90},\ref{aff91}}
\and Q.~Le~Boulc'h\inst{\ref{aff46}}
\and A.~M.~C.~Le~Brun\orcid{0000-0002-0936-4594}\inst{\ref{aff92}}
\and P.~Liebing\inst{\ref{aff93}}
\and V.~Lindholm\orcid{0000-0003-2317-5471}\inst{\ref{aff90},\ref{aff91}}
\and E.~Maiorano\orcid{0000-0003-2593-4355}\inst{\ref{aff9}}
\and O.~Mansutti\orcid{0000-0001-5758-4658}\inst{\ref{aff42}}
\and S.~Marcin\inst{\ref{aff94}}
\and O.~Marggraf\orcid{0000-0001-7242-3852}\inst{\ref{aff95}}
\and K.~Markovic\orcid{0000-0001-6764-073X}\inst{\ref{aff22}}
\and M.~Martinelli\orcid{0000-0002-6943-7732}\inst{\ref{aff51},\ref{aff52}}
\and N.~Martinet\orcid{0000-0003-2786-7790}\inst{\ref{aff3}}
\and F.~Marulli\orcid{0000-0002-8850-0303}\inst{\ref{aff47},\ref{aff9},\ref{aff34}}
\and R.~Massey\orcid{0000-0002-6085-3780}\inst{\ref{aff96}}
\and S.~Maurogordato\inst{\ref{aff97}}
\and H.~J.~McCracken\orcid{0000-0002-9489-7765}\inst{\ref{aff31}}
\and S.~Mei\orcid{0000-0002-2849-559X}\inst{\ref{aff98},\ref{aff99}}
\and M.~Melchior\inst{\ref{aff100}}
\and M.~Meneghetti\orcid{0000-0003-1225-7084}\inst{\ref{aff9},\ref{aff34}}
\and E.~Merlin\orcid{0000-0001-6870-8900}\inst{\ref{aff51}}
\and G.~Meylan\inst{\ref{aff101}}
\and A.~Mora\orcid{0000-0002-1922-8529}\inst{\ref{aff102}}
\and L.~Moscardini\orcid{0000-0002-3473-6716}\inst{\ref{aff47},\ref{aff9},\ref{aff34}}
\and R.~Nakajima\orcid{0009-0009-1213-7040}\inst{\ref{aff95}}
\and C.~Neissner\orcid{0000-0001-8524-4968}\inst{\ref{aff14},\ref{aff39}}
\and R.~C.~Nichol\orcid{0000-0003-0939-6518}\inst{\ref{aff55}}
\and S.-M.~Niemi\orcid{0009-0005-0247-0086}\inst{\ref{aff27}}
\and J.~W.~Nightingale\orcid{0000-0002-8987-7401}\inst{\ref{aff103}}
\and S.~Paltani\orcid{0000-0002-8108-9179}\inst{\ref{aff83}}
\and F.~Pasian\orcid{0000-0002-4869-3227}\inst{\ref{aff42}}
\and K.~Pedersen\inst{\ref{aff104}}
\and V.~Pettorino\inst{\ref{aff27}}
\and S.~Pires\orcid{0000-0002-0249-2104}\inst{\ref{aff21}}
\and M.~Poncet\inst{\ref{aff2}}
\and L.~A.~Popa\inst{\ref{aff105}}
\and L.~Pozzetti\orcid{0000-0001-7085-0412}\inst{\ref{aff9}}
\and F.~Raison\orcid{0000-0002-7819-6918}\inst{\ref{aff4}}
\and R.~Rebolo\orcid{0000-0003-3767-7085}\inst{\ref{aff73},\ref{aff106},\ref{aff107}}
\and J.~Rhodes\orcid{0000-0002-4485-8549}\inst{\ref{aff22}}
\and G.~Riccio\inst{\ref{aff63}}
\and E.~Romelli\orcid{0000-0003-3069-9222}\inst{\ref{aff42}}
\and M.~Roncarelli\orcid{0000-0001-9587-7822}\inst{\ref{aff9}}
\and E.~Rossetti\orcid{0000-0003-0238-4047}\inst{\ref{aff59}}
\and R.~Saglia\orcid{0000-0003-0378-7032}\inst{\ref{aff5},\ref{aff4}}
\and Z.~Sakr\orcid{0000-0002-4823-3757}\inst{\ref{aff108},\ref{aff109},\ref{aff110}}
\and A.~G.~S\'anchez\orcid{0000-0003-1198-831X}\inst{\ref{aff4}}
\and D.~Sapone\orcid{0000-0001-7089-4503}\inst{\ref{aff111}}
\and B.~Sartoris\orcid{0000-0003-1337-5269}\inst{\ref{aff5},\ref{aff42}}
\and J.~A.~Schewtschenko\orcid{0000-0002-4913-6393}\inst{\ref{aff74}}
\and P.~Schneider\orcid{0000-0001-8561-2679}\inst{\ref{aff95}}
\and T.~Schrabback\orcid{0000-0002-6987-7834}\inst{\ref{aff112}}
\and E.~Sefusatti\orcid{0000-0003-0473-1567}\inst{\ref{aff42},\ref{aff56},\ref{aff57}}
\and P.~Simon\inst{\ref{aff95}}
\and J.~Steinwagner\orcid{0000-0001-7443-1047}\inst{\ref{aff4}}
\and P.~Tallada-Cresp\'{i}\orcid{0000-0002-1336-8328}\inst{\ref{aff40},\ref{aff39}}
\and D.~Tavagnacco\orcid{0000-0001-7475-9894}\inst{\ref{aff42}}
\and A.~N.~Taylor\inst{\ref{aff74}}
\and H.~I.~Teplitz\orcid{0000-0002-7064-5424}\inst{\ref{aff37}}
\and I.~Tereno\orcid{0000-0002-4537-6218}\inst{\ref{aff81},\ref{aff113}}
\and S.~Toft\orcid{0000-0003-3631-7176}\inst{\ref{aff114},\ref{aff115}}
\and I.~Tutusaus\orcid{0000-0002-3199-0399}\inst{\ref{aff109}}
\and J.~Valiviita\orcid{0000-0001-6225-3693}\inst{\ref{aff90},\ref{aff91}}
\and T.~Vassallo\orcid{0000-0001-6512-6358}\inst{\ref{aff5},\ref{aff42}}
\and G.~Verdoes~Kleijn\orcid{0000-0001-5803-2580}\inst{\ref{aff28}}
\and A.~Veropalumbo\orcid{0000-0003-2387-1194}\inst{\ref{aff36},\ref{aff61},\ref{aff60}}
\and J.~Weller\orcid{0000-0002-8282-2010}\inst{\ref{aff5},\ref{aff4}}
\and A.~Zacchei\orcid{0000-0003-0396-1192}\inst{\ref{aff42},\ref{aff56}}
\and G.~Zamorani\orcid{0000-0002-2318-301X}\inst{\ref{aff9}}
\and F.~M.~Zerbi\inst{\ref{aff36}}
\and I.~A.~Zinchenko\orcid{0000-0002-2944-2449}\inst{\ref{aff5}}
\and E.~Zucca\orcid{0000-0002-5845-8132}\inst{\ref{aff9}}
\and V.~Allevato\orcid{0000-0001-7232-5152}\inst{\ref{aff63}}
\and M.~Ballardini\orcid{0000-0003-4481-3559}\inst{\ref{aff116},\ref{aff117},\ref{aff9}}
\and M.~Bolzonella\orcid{0000-0003-3278-4607}\inst{\ref{aff9}}
\and E.~Bozzo\orcid{0000-0002-8201-1525}\inst{\ref{aff83}}
\and C.~Burigana\orcid{0000-0002-3005-5796}\inst{\ref{aff118},\ref{aff16}}
\and R.~Cabanac\orcid{0000-0001-6679-2600}\inst{\ref{aff109}}
\and A.~Cappi\inst{\ref{aff9},\ref{aff97}}
\and D.~Di~Ferdinando\inst{\ref{aff34}}
\and J.~A.~Escartin~Vigo\inst{\ref{aff4}}
\and L.~Gabarra\orcid{0000-0002-8486-8856}\inst{\ref{aff119}}
\and W.~G.~Hartley\inst{\ref{aff83}}
\and J.~Mart\'{i}n-Fleitas\orcid{0000-0002-8594-569X}\inst{\ref{aff102}}
\and S.~Matthew\orcid{0000-0001-8448-1697}\inst{\ref{aff74}}
\and N.~Mauri\orcid{0000-0001-8196-1548}\inst{\ref{aff72},\ref{aff34}}
\and R.~B.~Metcalf\orcid{0000-0003-3167-2574}\inst{\ref{aff47},\ref{aff9}}
\and A.~Pezzotta\orcid{0000-0003-0726-2268}\inst{\ref{aff120},\ref{aff4}}
\and M.~P\"ontinen\orcid{0000-0001-5442-2530}\inst{\ref{aff90}}
\and C.~Porciani\orcid{0000-0002-7797-2508}\inst{\ref{aff95}}
\and I.~Risso\orcid{0000-0003-2525-7761}\inst{\ref{aff121}}
\and V.~Scottez\inst{\ref{aff30},\ref{aff122}}
\and M.~Sereno\orcid{0000-0003-0302-0325}\inst{\ref{aff9},\ref{aff34}}
\and M.~Tenti\orcid{0000-0002-4254-5901}\inst{\ref{aff34}}
\and M.~Viel\orcid{0000-0002-2642-5707}\inst{\ref{aff56},\ref{aff42},\ref{aff58},\ref{aff57},\ref{aff123}}
\and M.~Wiesmann\orcid{0009-0000-8199-5860}\inst{\ref{aff19}}
\and Y.~Akrami\orcid{0000-0002-2407-7956}\inst{\ref{aff124},\ref{aff125}}
\and I.~T.~Andika\orcid{0000-0001-6102-9526}\inst{\ref{aff126},\ref{aff127}}
\and S.~Anselmi\orcid{0000-0002-3579-9583}\inst{\ref{aff7},\ref{aff8},\ref{aff128}}
\and M.~Archidiacono\orcid{0000-0003-4952-9012}\inst{\ref{aff44},\ref{aff45}}
\and F.~Atrio-Barandela\orcid{0000-0002-2130-2513}\inst{\ref{aff129}}
\and D.~Bertacca\orcid{0000-0002-2490-7139}\inst{\ref{aff8},\ref{aff12},\ref{aff7}}
\and M.~Bethermin\orcid{0000-0002-3915-2015}\inst{\ref{aff130}}
\and A.~Blanchard\orcid{0000-0001-8555-9003}\inst{\ref{aff109}}
\and L.~Blot\orcid{0000-0002-9622-7167}\inst{\ref{aff131},\ref{aff92}}
\and S.~Borgani\orcid{0000-0001-6151-6439}\inst{\ref{aff132},\ref{aff56},\ref{aff42},\ref{aff57},\ref{aff123}}
\and M.~L.~Brown\orcid{0000-0002-0370-8077}\inst{\ref{aff75}}
\and S.~Bruton\orcid{0000-0002-6503-5218}\inst{\ref{aff53}}
\and A.~Calabro\orcid{0000-0003-2536-1614}\inst{\ref{aff51}}
\and B.~Camacho~Quevedo\orcid{0000-0002-8789-4232}\inst{\ref{aff18},\ref{aff17}}
\and F.~Caro\inst{\ref{aff51}}
\and C.~S.~Carvalho\inst{\ref{aff113}}
\and T.~Castro\orcid{0000-0002-6292-3228}\inst{\ref{aff42},\ref{aff57},\ref{aff56},\ref{aff123}}
\and Y.~Charles\inst{\ref{aff3}}
\and F.~Cogato\orcid{0000-0003-4632-6113}\inst{\ref{aff47},\ref{aff9}}
\and S.~Conseil\orcid{0000-0002-3657-4191}\inst{\ref{aff11}}
\and A.~R.~Cooray\orcid{0000-0002-3892-0190}\inst{\ref{aff133}}
\and O.~Cucciati\orcid{0000-0002-9336-7551}\inst{\ref{aff9}}
\and S.~Davini\orcid{0000-0003-3269-1718}\inst{\ref{aff61}}
\and F.~De~Paolis\orcid{0000-0001-6460-7563}\inst{\ref{aff134},\ref{aff135},\ref{aff136}}
\and G.~Desprez\orcid{0000-0001-8325-1742}\inst{\ref{aff28}}
\and A.~D\'iaz-S\'anchez\orcid{0000-0003-0748-4768}\inst{\ref{aff137}}
\and J.~J.~Diaz\inst{\ref{aff73}}
\and S.~Di~Domizio\orcid{0000-0003-2863-5895}\inst{\ref{aff60},\ref{aff61}}
\and J.~M.~Diego\orcid{0000-0001-9065-3926}\inst{\ref{aff138}}
\and P.~Dimauro\orcid{0000-0001-7399-2854}\inst{\ref{aff51},\ref{aff139}}
\and P.-A.~Duc\orcid{0000-0003-3343-6284}\inst{\ref{aff130}}
\and A.~Enia\orcid{0000-0002-0200-2857}\inst{\ref{aff59},\ref{aff9}}
\and Y.~Fang\inst{\ref{aff5}}
\and A.~M.~N.~Ferguson\inst{\ref{aff74}}
\and A.~G.~Ferrari\orcid{0009-0005-5266-4110}\inst{\ref{aff34}}
\and A.~Finoguenov\orcid{0000-0002-4606-5403}\inst{\ref{aff90}}
\and A.~Franco\orcid{0000-0002-4761-366X}\inst{\ref{aff135},\ref{aff134},\ref{aff136}}
\and K.~Ganga\orcid{0000-0001-8159-8208}\inst{\ref{aff98}}
\and J.~Garc\'ia-Bellido\orcid{0000-0002-9370-8360}\inst{\ref{aff124}}
\and T.~Gasparetto\orcid{0000-0002-7913-4866}\inst{\ref{aff42}}
\and V.~Gautard\inst{\ref{aff140}}
\and E.~Gaztanaga\orcid{0000-0001-9632-0815}\inst{\ref{aff17},\ref{aff18},\ref{aff141}}
\and F.~Giacomini\orcid{0000-0002-3129-2814}\inst{\ref{aff34}}
\and F.~Gianotti\orcid{0000-0003-4666-119X}\inst{\ref{aff9}}
\and G.~Gozaliasl\orcid{0000-0002-0236-919X}\inst{\ref{aff142},\ref{aff90}}
\and A.~Gregorio\orcid{0000-0003-4028-8785}\inst{\ref{aff132},\ref{aff42},\ref{aff57}}
\and M.~Guidi\orcid{0000-0001-9408-1101}\inst{\ref{aff59},\ref{aff9}}
\and C.~M.~Gutierrez\orcid{0000-0001-7854-783X}\inst{\ref{aff143}}
\and A.~Hall\orcid{0000-0002-3139-8651}\inst{\ref{aff74}}
\and S.~Hemmati\orcid{0000-0003-2226-5395}\inst{\ref{aff41}}
\and C.~Hern\'andez-Monteagudo\orcid{0000-0001-5471-9166}\inst{\ref{aff107},\ref{aff73}}
\and H.~Hildebrandt\orcid{0000-0002-9814-3338}\inst{\ref{aff144}}
\and J.~Hjorth\orcid{0000-0002-4571-2306}\inst{\ref{aff104}}
\and J.~J.~E.~Kajava\orcid{0000-0002-3010-8333}\inst{\ref{aff145},\ref{aff146}}
\and Y.~Kang\orcid{0009-0000-8588-7250}\inst{\ref{aff83}}
\and V.~Kansal\orcid{0000-0002-4008-6078}\inst{\ref{aff147},\ref{aff148}}
\and D.~Karagiannis\orcid{0000-0002-4927-0816}\inst{\ref{aff116},\ref{aff149}}
\and K.~Kiiveri\inst{\ref{aff88}}
\and C.~C.~Kirkpatrick\inst{\ref{aff88}}
\and S.~Kruk\orcid{0000-0001-8010-8879}\inst{\ref{aff26}}
\and J.~Le~Graet\orcid{0000-0001-6523-7971}\inst{\ref{aff1}}
\and L.~Legrand\orcid{0000-0003-0610-5252}\inst{\ref{aff150},\ref{aff151}}
\and M.~Lembo\orcid{0000-0002-5271-5070}\inst{\ref{aff116},\ref{aff117}}
\and F.~Lepori\orcid{0009-0000-5061-7138}\inst{\ref{aff152}}
\and G.~Leroy\orcid{0009-0004-2523-4425}\inst{\ref{aff153},\ref{aff96}}
\and G.~F.~Lesci\orcid{0000-0002-4607-2830}\inst{\ref{aff47},\ref{aff9}}
\and J.~Lesgourgues\orcid{0000-0001-7627-353X}\inst{\ref{aff69}}
\and L.~Leuzzi\orcid{0009-0006-4479-7017}\inst{\ref{aff47},\ref{aff9}}
\and T.~I.~Liaudat\orcid{0000-0002-9104-314X}\inst{\ref{aff154}}
\and S.~J.~Liu\orcid{0000-0001-7680-2139}\inst{\ref{aff84}}
\and A.~Loureiro\orcid{0000-0002-4371-0876}\inst{\ref{aff155},\ref{aff156}}
\and J.~Macias-Perez\orcid{0000-0002-5385-2763}\inst{\ref{aff157}}
\and G.~Maggio\orcid{0000-0003-4020-4836}\inst{\ref{aff42}}
\and M.~Magliocchetti\orcid{0000-0001-9158-4838}\inst{\ref{aff84}}
\and C.~Mancini\orcid{0000-0002-4297-0561}\inst{\ref{aff43}}
\and F.~Mannucci\orcid{0000-0002-4803-2381}\inst{\ref{aff158}}
\and R.~Maoli\orcid{0000-0002-6065-3025}\inst{\ref{aff159},\ref{aff51}}
\and C.~J.~A.~P.~Martins\orcid{0000-0002-4886-9261}\inst{\ref{aff160},\ref{aff64}}
\and L.~Maurin\orcid{0000-0002-8406-0857}\inst{\ref{aff54}}
\and C.~J.~R.~McPartland\orcid{0000-0003-0639-025X}\inst{\ref{aff24},\ref{aff115}}
\and M.~Miluzio\inst{\ref{aff26},\ref{aff161}}
\and P.~Monaco\orcid{0000-0003-2083-7564}\inst{\ref{aff132},\ref{aff42},\ref{aff57},\ref{aff56}}
\and A.~Montoro\orcid{0000-0003-4730-8590}\inst{\ref{aff17},\ref{aff18}}
\and C.~Moretti\orcid{0000-0003-3314-8936}\inst{\ref{aff58},\ref{aff123},\ref{aff42},\ref{aff56},\ref{aff57}}
\and S.~Nadathur\orcid{0000-0001-9070-3102}\inst{\ref{aff141}}
\and K.~Naidoo\orcid{0000-0002-9182-1802}\inst{\ref{aff141}}
\and A.~Navarro-Alsina\orcid{0000-0002-3173-2592}\inst{\ref{aff95}}
\and F.~Passalacqua\orcid{0000-0002-8606-4093}\inst{\ref{aff8},\ref{aff7}}
\and K.~Paterson\orcid{0000-0001-8340-3486}\inst{\ref{aff6}}
\and L.~Patrizii\inst{\ref{aff34}}
\and A.~Pisani\orcid{0000-0002-6146-4437}\inst{\ref{aff1},\ref{aff162}}
\and D.~Potter\orcid{0000-0002-0757-5195}\inst{\ref{aff152}}
\and S.~Quai\orcid{0000-0002-0449-8163}\inst{\ref{aff47},\ref{aff9}}
\and M.~Radovich\orcid{0000-0002-3585-866X}\inst{\ref{aff12}}
\and P.-F.~Rocci\inst{\ref{aff54}}
\and S.~Sacquegna\orcid{0000-0002-8433-6630}\inst{\ref{aff134},\ref{aff135},\ref{aff136}}
\and M.~Sahl\'en\orcid{0000-0003-0973-4804}\inst{\ref{aff163}}
\and D.~B.~Sanders\orcid{0000-0002-1233-9998}\inst{\ref{aff71}}
\and E.~Sarpa\orcid{0000-0002-1256-655X}\inst{\ref{aff58},\ref{aff123},\ref{aff57}}
\and A.~Schneider\orcid{0000-0001-7055-8104}\inst{\ref{aff152}}
\and D.~Sciotti\orcid{0009-0008-4519-2620}\inst{\ref{aff51},\ref{aff52}}
\and E.~Sellentin\inst{\ref{aff164},\ref{aff33}}
\and L.~C.~Smith\orcid{0000-0002-3259-2771}\inst{\ref{aff165}}
\and K.~Tanidis\orcid{0000-0001-9843-5130}\inst{\ref{aff119}}
\and C.~Tao\orcid{0000-0001-7961-8177}\inst{\ref{aff1}}
\and G.~Testera\inst{\ref{aff61}}
\and R.~Teyssier\orcid{0000-0001-7689-0933}\inst{\ref{aff162}}
\and S.~Tosi\orcid{0000-0002-7275-9193}\inst{\ref{aff60},\ref{aff61},\ref{aff36}}
\and A.~Troja\orcid{0000-0003-0239-4595}\inst{\ref{aff8},\ref{aff7}}
\and M.~Tucci\inst{\ref{aff83}}
\and C.~Valieri\inst{\ref{aff34}}
\and A.~Venhola\orcid{0000-0001-6071-4564}\inst{\ref{aff166}}
\and D.~Vergani\orcid{0000-0003-0898-2216}\inst{\ref{aff9}}
\and G.~Verza\orcid{0000-0002-1886-8348}\inst{\ref{aff167}}
\and N.~A.~Walton\orcid{0000-0003-3983-8778}\inst{\ref{aff165}}
\and L.~Zalesky\orcid{0000-0001-5680-2326}\inst{\ref{aff71}}}

\institute{Aix-Marseille Universit\'e, CNRS/IN2P3, CPPM, Marseille, France\label{aff1}
\and
Centre National d'Etudes Spatiales -- Centre spatial de Toulouse, 18 avenue Edouard Belin, 31401 Toulouse Cedex 9, France\label{aff2}
\and
Aix-Marseille Universit\'e, CNRS, CNES, LAM, Marseille, France\label{aff3}
\and
Max Planck Institute for Extraterrestrial Physics, Giessenbachstr. 1, 85748 Garching, Germany\label{aff4}
\and
Universit\"ats-Sternwarte M\"unchen, Fakult\"at f\"ur Physik, Ludwig-Maximilians-Universit\"at M\"unchen, Scheinerstrasse 1, 81679 M\"unchen, Germany\label{aff5}
\and
Max-Planck-Institut f\"ur Astronomie, K\"onigstuhl 17, 69117 Heidelberg, Germany\label{aff6}
\and
INFN-Padova, Via Marzolo 8, 35131 Padova, Italy\label{aff7}
\and
Dipartimento di Fisica e Astronomia "G. Galilei", Universit\`a di Padova, Via Marzolo 8, 35131 Padova, Italy\label{aff8}
\and
INAF-Osservatorio di Astrofisica e Scienza dello Spazio di Bologna, Via Piero Gobetti 93/3, 40129 Bologna, Italy\label{aff9}
\and
INAF-Osservatorio Astrofisico di Torino, Via Osservatorio 20, 10025 Pino Torinese (TO), Italy\label{aff10}
\and
Universit\'e Claude Bernard Lyon 1, CNRS/IN2P3, IP2I Lyon, UMR 5822, Villeurbanne, F-69100, France\label{aff11}
\and
INAF-Osservatorio Astronomico di Padova, Via dell'Osservatorio 5, 35122 Padova, Italy\label{aff12}
\and
Felix Hormuth Engineering, Goethestr. 17, 69181 Leimen, Germany\label{aff13}
\and
Institut de F\'{i}sica d'Altes Energies (IFAE), The Barcelona Institute of Science and Technology, Campus UAB, 08193 Bellaterra (Barcelona), Spain\label{aff14}
\and
Universidad Polit\'ecnica de Cartagena, Departamento de Electr\'onica y Tecnolog\'ia de Computadoras,  Plaza del Hospital 1, 30202 Cartagena, Spain\label{aff15}
\and
INFN-Bologna, Via Irnerio 46, 40126 Bologna, Italy\label{aff16}
\and
Institute of Space Sciences (ICE, CSIC), Campus UAB, Carrer de Can Magrans, s/n, 08193 Barcelona, Spain\label{aff17}
\and
Institut d'Estudis Espacials de Catalunya (IEEC),  Edifici RDIT, Campus UPC, 08860 Castelldefels, Barcelona, Spain\label{aff18}
\and
Institute of Theoretical Astrophysics, University of Oslo, P.O. Box 1029 Blindern, 0315 Oslo, Norway\label{aff19}
\and
CEA-Saclay, DRF/IRFU, departement d'ingenierie des systemes, bat472, 91191 Gif sur Yvette cedex, France\label{aff20}
\and
Universit\'e Paris-Saclay, Universit\'e Paris Cit\'e, CEA, CNRS, AIM, 91191, Gif-sur-Yvette, France\label{aff21}
\and
Jet Propulsion Laboratory, California Institute of Technology, 4800 Oak Grove Drive, Pasadena, CA, 91109, USA\label{aff22}
\and
Technical University of Denmark, Elektrovej 327, 2800 Kgs. Lyngby, Denmark\label{aff23}
\and
Cosmic Dawn Center (DAWN), Denmark\label{aff24}
\and
NASA Goddard Space Flight Center, Greenbelt, MD 20771, USA\label{aff25}
\and
ESAC/ESA, Camino Bajo del Castillo, s/n., Urb. Villafranca del Castillo, 28692 Villanueva de la Ca\~nada, Madrid, Spain\label{aff26}
\and
European Space Agency/ESTEC, Keplerlaan 1, 2201 AZ Noordwijk, The Netherlands\label{aff27}
\and
Kapteyn Astronomical Institute, University of Groningen, PO Box 800, 9700 AV Groningen, The Netherlands\label{aff28}
\and
SKA Observatory, Jodrell Bank, Lower Withington, Macclesfield, Cheshire SK11 9FT, UK\label{aff29}
\and
Institut d'Astrophysique de Paris, 98bis Boulevard Arago, 75014, Paris, France\label{aff30}
\and
Institut d'Astrophysique de Paris, UMR 7095, CNRS, and Sorbonne Universit\'e, 98 bis boulevard Arago, 75014 Paris, France\label{aff31}
\and
Space Science Data Center, Italian Space Agency, via del Politecnico snc, 00133 Roma, Italy\label{aff32}
\and
Leiden Observatory, Leiden University, Einsteinweg 55, 2333 CC Leiden, The Netherlands\label{aff33}
\and
INFN-Sezione di Bologna, Viale Berti Pichat 6/2, 40127 Bologna, Italy\label{aff34}
\and
Carnegie Observatories, Pasadena, CA 91101, USA\label{aff35}
\and
INAF-Osservatorio Astronomico di Brera, Via Brera 28, 20122 Milano, Italy\label{aff36}
\and
Infrared Processing and Analysis Center, California Institute of Technology, Pasadena, CA 91125, USA\label{aff37}
\and
University of California, Los Angeles, CA 90095-1562, USA\label{aff38}
\and
Port d'Informaci\'{o} Cient\'{i}fica, Campus UAB, C. Albareda s/n, 08193 Bellaterra (Barcelona), Spain\label{aff39}
\and
Centro de Investigaciones Energ\'eticas, Medioambientales y Tecnol\'ogicas (CIEMAT), Avenida Complutense 40, 28040 Madrid, Spain\label{aff40}
\and
Caltech/IPAC, 1200 E. California Blvd., Pasadena, CA 91125, USA\label{aff41}
\and
INAF-Osservatorio Astronomico di Trieste, Via G. B. Tiepolo 11, 34143 Trieste, Italy\label{aff42}
\and
INAF-IASF Milano, Via Alfonso Corti 12, 20133 Milano, Italy\label{aff43}
\and
Dipartimento di Fisica "Aldo Pontremoli", Universit\`a degli Studi di Milano, Via Celoria 16, 20133 Milano, Italy\label{aff44}
\and
INFN-Sezione di Milano, Via Celoria 16, 20133 Milano, Italy\label{aff45}
\and
Centre de Calcul de l'IN2P3/CNRS, 21 avenue Pierre de Coubertin 69627 Villeurbanne Cedex, France\label{aff46}
\and
Dipartimento di Fisica e Astronomia "Augusto Righi" - Alma Mater Studiorum Universit\`a di Bologna, via Piero Gobetti 93/2, 40129 Bologna, Italy\label{aff47}
\and
Waterloo Centre for Astrophysics, University of Waterloo, Waterloo, Ontario N2L 3G1, Canada\label{aff48}
\and
Department of Physics and Astronomy, University of Waterloo, Waterloo, Ontario N2L 3G1, Canada\label{aff49}
\and
Perimeter Institute for Theoretical Physics, Waterloo, Ontario N2L 2Y5, Canada\label{aff50}
\and
INAF-Osservatorio Astronomico di Roma, Via Frascati 33, 00078 Monteporzio Catone, Italy\label{aff51}
\and
INFN-Sezione di Roma, Piazzale Aldo Moro, 2 - c/o Dipartimento di Fisica, Edificio G. Marconi, 00185 Roma, Italy\label{aff52}
\and
California Institute of Technology, 1200 E California Blvd, Pasadena, CA 91125, USA\label{aff53}
\and
Universit\'e Paris-Saclay, CNRS, Institut d'astrophysique spatiale, 91405, Orsay, France\label{aff54}
\and
School of Mathematics and Physics, University of Surrey, Guildford, Surrey, GU2 7XH, UK\label{aff55}
\and
IFPU, Institute for Fundamental Physics of the Universe, via Beirut 2, 34151 Trieste, Italy\label{aff56}
\and
INFN, Sezione di Trieste, Via Valerio 2, 34127 Trieste TS, Italy\label{aff57}
\and
SISSA, International School for Advanced Studies, Via Bonomea 265, 34136 Trieste TS, Italy\label{aff58}
\and
Dipartimento di Fisica e Astronomia, Universit\`a di Bologna, Via Gobetti 93/2, 40129 Bologna, Italy\label{aff59}
\and
Dipartimento di Fisica, Universit\`a di Genova, Via Dodecaneso 33, 16146, Genova, Italy\label{aff60}
\and
INFN-Sezione di Genova, Via Dodecaneso 33, 16146, Genova, Italy\label{aff61}
\and
Department of Physics "E. Pancini", University Federico II, Via Cinthia 6, 80126, Napoli, Italy\label{aff62}
\and
INAF-Osservatorio Astronomico di Capodimonte, Via Moiariello 16, 80131 Napoli, Italy\label{aff63}
\and
Instituto de Astrof\'isica e Ci\^encias do Espa\c{c}o, Universidade do Porto, CAUP, Rua das Estrelas, PT4150-762 Porto, Portugal\label{aff64}
\and
Faculdade de Ci\^encias da Universidade do Porto, Rua do Campo de Alegre, 4150-007 Porto, Portugal\label{aff65}
\and
Dipartimento di Fisica, Universit\`a degli Studi di Torino, Via P. Giuria 1, 10125 Torino, Italy\label{aff66}
\and
INFN-Sezione di Torino, Via P. Giuria 1, 10125 Torino, Italy\label{aff67}
\and
Institute Lorentz, Leiden University, Niels Bohrweg 2, 2333 CA Leiden, The Netherlands\label{aff68}
\and
Institute for Theoretical Particle Physics and Cosmology (TTK), RWTH Aachen University, 52056 Aachen, Germany\label{aff69}
\and
INFN section of Naples, Via Cinthia 6, 80126, Napoli, Italy\label{aff70}
\and
Institute for Astronomy, University of Hawaii, 2680 Woodlawn Drive, Honolulu, HI 96822, USA\label{aff71}
\and
Dipartimento di Fisica e Astronomia "Augusto Righi" - Alma Mater Studiorum Universit\`a di Bologna, Viale Berti Pichat 6/2, 40127 Bologna, Italy\label{aff72}
\and
Instituto de Astrof\'{\i}sica de Canarias, V\'{\i}a L\'actea, 38205 La Laguna, Tenerife, Spain\label{aff73}
\and
Institute for Astronomy, University of Edinburgh, Royal Observatory, Blackford Hill, Edinburgh EH9 3HJ, UK\label{aff74}
\and
Jodrell Bank Centre for Astrophysics, Department of Physics and Astronomy, University of Manchester, Oxford Road, Manchester M13 9PL, UK\label{aff75}
\and
European Space Agency/ESRIN, Largo Galileo Galilei 1, 00044 Frascati, Roma, Italy\label{aff76}
\and
Institut de Ci\`{e}ncies del Cosmos (ICCUB), Universitat de Barcelona (IEEC-UB), Mart\'{i} i Franqu\`{e}s 1, 08028 Barcelona, Spain\label{aff77}
\and
Instituci\'o Catalana de Recerca i Estudis Avan\c{c}ats (ICREA), Passeig de Llu\'{\i}s Companys 23, 08010 Barcelona, Spain\label{aff78}
\and
UCB Lyon 1, CNRS/IN2P3, IUF, IP2I Lyon, 4 rue Enrico Fermi, 69622 Villeurbanne, France\label{aff79}
\and
Canada-France-Hawaii Telescope, 65-1238 Mamalahoa Hwy, Kamuela, HI 96743, USA\label{aff80}
\and
Departamento de F\'isica, Faculdade de Ci\^encias, Universidade de Lisboa, Edif\'icio C8, Campo Grande, PT1749-016 Lisboa, Portugal\label{aff81}
\and
Instituto de Astrof\'isica e Ci\^encias do Espa\c{c}o, Faculdade de Ci\^encias, Universidade de Lisboa, Campo Grande, 1749-016 Lisboa, Portugal\label{aff82}
\and
Department of Astronomy, University of Geneva, ch. d'Ecogia 16, 1290 Versoix, Switzerland\label{aff83}
\and
INAF-Istituto di Astrofisica e Planetologia Spaziali, via del Fosso del Cavaliere, 100, 00100 Roma, Italy\label{aff84}
\and
School of Physics, HH Wills Physics Laboratory, University of Bristol, Tyndall Avenue, Bristol, BS8 1TL, UK\label{aff85}
\and
FRACTAL S.L.N.E., calle Tulip\'an 2, Portal 13 1A, 28231, Las Rozas de Madrid, Spain\label{aff86}
\and
Department of Physics, Lancaster University, Lancaster, LA1 4YB, UK\label{aff87}
\and
Department of Physics and Helsinki Institute of Physics, Gustaf H\"allstr\"omin katu 2, 00014 University of Helsinki, Finland\label{aff88}
\and
Universit\'e de Gen\`eve, D\'epartement de Physique Th\'eorique and Centre for Astroparticle Physics, 24 quai Ernest-Ansermet, CH-1211 Gen\`eve 4, Switzerland\label{aff89}
\and
Department of Physics, P.O. Box 64, 00014 University of Helsinki, Finland\label{aff90}
\and
Helsinki Institute of Physics, Gustaf H{\"a}llstr{\"o}min katu 2, University of Helsinki, Helsinki, Finland\label{aff91}
\and
Laboratoire d'etude de l'Univers et des phenomenes eXtremes, Observatoire de Paris, Universit\'e PSL, Sorbonne Universit\'e, CNRS, 92190 Meudon, France\label{aff92}
\and
Mullard Space Science Laboratory, University College London, Holmbury St Mary, Dorking, Surrey RH5 6NT, UK\label{aff93}
\and
University of Applied Sciences and Arts of Northwestern Switzerland, School of Computer Science, 5210 Windisch, Switzerland\label{aff94}
\and
Universit\"at Bonn, Argelander-Institut f\"ur Astronomie, Auf dem H\"ugel 71, 53121 Bonn, Germany\label{aff95}
\and
Department of Physics, Institute for Computational Cosmology, Durham University, South Road, Durham, DH1 3LE, UK\label{aff96}
\and
Universit\'e C\^{o}te d'Azur, Observatoire de la C\^{o}te d'Azur, CNRS, Laboratoire Lagrange, Bd de l'Observatoire, CS 34229, 06304 Nice cedex 4, France\label{aff97}
\and
Universit\'e Paris Cit\'e, CNRS, Astroparticule et Cosmologie, 75013 Paris, France\label{aff98}
\and
CNRS-UCB International Research Laboratory, Centre Pierre Bin\'etruy, IRL2007, CPB-IN2P3, Berkeley, USA\label{aff99}
\and
University of Applied Sciences and Arts of Northwestern Switzerland, School of Engineering, 5210 Windisch, Switzerland\label{aff100}
\and
Institute of Physics, Laboratory of Astrophysics, Ecole Polytechnique F\'ed\'erale de Lausanne (EPFL), Observatoire de Sauverny, 1290 Versoix, Switzerland\label{aff101}
\and
Aurora Technology for European Space Agency (ESA), Camino bajo del Castillo, s/n, Urbanizacion Villafranca del Castillo, Villanueva de la Ca\~nada, 28692 Madrid, Spain\label{aff102}
\and
School of Mathematics, Statistics and Physics, Newcastle University, Herschel Building, Newcastle-upon-Tyne, NE1 7RU, UK\label{aff103}
\and
DARK, Niels Bohr Institute, University of Copenhagen, Jagtvej 155, 2200 Copenhagen, Denmark\label{aff104}
\and
Institute of Space Science, Str. Atomistilor, nr. 409 M\u{a}gurele, Ilfov, 077125, Romania\label{aff105}
\and
Consejo Superior de Investigaciones Cientificas, Calle Serrano 117, 28006 Madrid, Spain\label{aff106}
\and
Universidad de La Laguna, Departamento de Astrof\'{\i}sica, 38206 La Laguna, Tenerife, Spain\label{aff107}
\and
Institut f\"ur Theoretische Physik, University of Heidelberg, Philosophenweg 16, 69120 Heidelberg, Germany\label{aff108}
\and
Institut de Recherche en Astrophysique et Plan\'etologie (IRAP), Universit\'e de Toulouse, CNRS, UPS, CNES, 14 Av. Edouard Belin, 31400 Toulouse, France\label{aff109}
\and
Universit\'e St Joseph; Faculty of Sciences, Beirut, Lebanon\label{aff110}
\and
Departamento de F\'isica, FCFM, Universidad de Chile, Blanco Encalada 2008, Santiago, Chile\label{aff111}
\and
Universit\"at Innsbruck, Institut f\"ur Astro- und Teilchenphysik, Technikerstr. 25/8, 6020 Innsbruck, Austria\label{aff112}
\and
Instituto de Astrof\'isica e Ci\^encias do Espa\c{c}o, Faculdade de Ci\^encias, Universidade de Lisboa, Tapada da Ajuda, 1349-018 Lisboa, Portugal\label{aff113}
\and
Cosmic Dawn Center (DAWN)\label{aff114}
\and
Niels Bohr Institute, University of Copenhagen, Jagtvej 128, 2200 Copenhagen, Denmark\label{aff115}
\and
Dipartimento di Fisica e Scienze della Terra, Universit\`a degli Studi di Ferrara, Via Giuseppe Saragat 1, 44122 Ferrara, Italy\label{aff116}
\and
Istituto Nazionale di Fisica Nucleare, Sezione di Ferrara, Via Giuseppe Saragat 1, 44122 Ferrara, Italy\label{aff117}
\and
INAF, Istituto di Radioastronomia, Via Piero Gobetti 101, 40129 Bologna, Italy\label{aff118}
\and
Department of Physics, Oxford University, Keble Road, Oxford OX1 3RH, UK\label{aff119}
\and
INAF - Osservatorio Astronomico di Brera, via Emilio Bianchi 46, 23807 Merate, Italy\label{aff120}
\and
INAF-Osservatorio Astronomico di Brera, Via Brera 28, 20122 Milano, Italy, and INFN-Sezione di Genova, Via Dodecaneso 33, 16146, Genova, Italy\label{aff121}
\and
ICL, Junia, Universit\'e Catholique de Lille, LITL, 59000 Lille, France\label{aff122}
\and
ICSC - Centro Nazionale di Ricerca in High Performance Computing, Big Data e Quantum Computing, Via Magnanelli 2, Bologna, Italy\label{aff123}
\and
Instituto de F\'isica Te\'orica UAM-CSIC, Campus de Cantoblanco, 28049 Madrid, Spain\label{aff124}
\and
CERCA/ISO, Department of Physics, Case Western Reserve University, 10900 Euclid Avenue, Cleveland, OH 44106, USA\label{aff125}
\and
Technical University of Munich, TUM School of Natural Sciences, Physics Department, James-Franck-Str.~1, 85748 Garching, Germany\label{aff126}
\and
Max-Planck-Institut f\"ur Astrophysik, Karl-Schwarzschild-Str.~1, 85748 Garching, Germany\label{aff127}
\and
Laboratoire Univers et Th\'eorie, Observatoire de Paris, Universit\'e PSL, Universit\'e Paris Cit\'e, CNRS, 92190 Meudon, France\label{aff128}
\and
Departamento de F{\'\i}sica Fundamental. Universidad de Salamanca. Plaza de la Merced s/n. 37008 Salamanca, Spain\label{aff129}
\and
Universit\'e de Strasbourg, CNRS, Observatoire astronomique de Strasbourg, UMR 7550, 67000 Strasbourg, France\label{aff130}
\and
Center for Data-Driven Discovery, Kavli IPMU (WPI), UTIAS, The University of Tokyo, Kashiwa, Chiba 277-8583, Japan\label{aff131}
\and
Dipartimento di Fisica - Sezione di Astronomia, Universit\`a di Trieste, Via Tiepolo 11, 34131 Trieste, Italy\label{aff132}
\and
Department of Physics \& Astronomy, University of California Irvine, Irvine CA 92697, USA\label{aff133}
\and
Department of Mathematics and Physics E. De Giorgi, University of Salento, Via per Arnesano, CP-I93, 73100, Lecce, Italy\label{aff134}
\and
INFN, Sezione di Lecce, Via per Arnesano, CP-193, 73100, Lecce, Italy\label{aff135}
\and
INAF-Sezione di Lecce, c/o Dipartimento Matematica e Fisica, Via per Arnesano, 73100, Lecce, Italy\label{aff136}
\and
Departamento F\'isica Aplicada, Universidad Polit\'ecnica de Cartagena, Campus Muralla del Mar, 30202 Cartagena, Murcia, Spain\label{aff137}
\and
Instituto de F\'isica de Cantabria, Edificio Juan Jord\'a, Avenida de los Castros, 39005 Santander, Spain\label{aff138}
\and
Observatorio Nacional, Rua General Jose Cristino, 77-Bairro Imperial de Sao Cristovao, Rio de Janeiro, 20921-400, Brazil\label{aff139}
\and
CEA Saclay, DFR/IRFU, Service d'Astrophysique, Bat. 709, 91191 Gif-sur-Yvette, France\label{aff140}
\and
Institute of Cosmology and Gravitation, University of Portsmouth, Portsmouth PO1 3FX, UK\label{aff141}
\and
Department of Computer Science, Aalto University, PO Box 15400, Espoo, FI-00 076, Finland\label{aff142}
\and
Instituto de Astrof\'\i sica de Canarias, c/ Via Lactea s/n, La Laguna 38200, Spain. Departamento de Astrof\'\i sica de la Universidad de La Laguna, Avda. Francisco Sanchez, La Laguna, 38200, Spain\label{aff143}
\and
Ruhr University Bochum, Faculty of Physics and Astronomy, Astronomical Institute (AIRUB), German Centre for Cosmological Lensing (GCCL), 44780 Bochum, Germany\label{aff144}
\and
Department of Physics and Astronomy, Vesilinnantie 5, 20014 University of Turku, Finland\label{aff145}
\and
Serco for European Space Agency (ESA), Camino bajo del Castillo, s/n, Urbanizacion Villafranca del Castillo, Villanueva de la Ca\~nada, 28692 Madrid, Spain\label{aff146}
\and
ARC Centre of Excellence for Dark Matter Particle Physics, Melbourne, Australia\label{aff147}
\and
Centre for Astrophysics \& Supercomputing, Swinburne University of Technology,  Hawthorn, Victoria 3122, Australia\label{aff148}
\and
Department of Physics and Astronomy, University of the Western Cape, Bellville, Cape Town, 7535, South Africa\label{aff149}
\and
DAMTP, Centre for Mathematical Sciences, Wilberforce Road, Cambridge CB3 0WA, UK\label{aff150}
\and
Kavli Institute for Cosmology Cambridge, Madingley Road, Cambridge, CB3 0HA, UK\label{aff151}
\and
Department of Astrophysics, University of Zurich, Winterthurerstrasse 190, 8057 Zurich, Switzerland\label{aff152}
\and
Department of Physics, Centre for Extragalactic Astronomy, Durham University, South Road, Durham, DH1 3LE, UK\label{aff153}
\and
IRFU, CEA, Universit\'e Paris-Saclay 91191 Gif-sur-Yvette Cedex, France\label{aff154}
\and
Oskar Klein Centre for Cosmoparticle Physics, Department of Physics, Stockholm University, Stockholm, SE-106 91, Sweden\label{aff155}
\and
Astrophysics Group, Blackett Laboratory, Imperial College London, London SW7 2AZ, UK\label{aff156}
\and
Univ. Grenoble Alpes, CNRS, Grenoble INP, LPSC-IN2P3, 53, Avenue des Martyrs, 38000, Grenoble, France\label{aff157}
\and
INAF-Osservatorio Astrofisico di Arcetri, Largo E. Fermi 5, 50125, Firenze, Italy\label{aff158}
\and
Dipartimento di Fisica, Sapienza Universit\`a di Roma, Piazzale Aldo Moro 2, 00185 Roma, Italy\label{aff159}
\and
Centro de Astrof\'{\i}sica da Universidade do Porto, Rua das Estrelas, 4150-762 Porto, Portugal\label{aff160}
\and
HE Space for European Space Agency (ESA), Camino bajo del Castillo, s/n, Urbanizacion Villafranca del Castillo, Villanueva de la Ca\~nada, 28692 Madrid, Spain\label{aff161}
\and
Department of Astrophysical Sciences, Peyton Hall, Princeton University, Princeton, NJ 08544, USA\label{aff162}
\and
Theoretical astrophysics, Department of Physics and Astronomy, Uppsala University, Box 516, 751 37 Uppsala, Sweden\label{aff163}
\and
Mathematical Institute, University of Leiden, Einsteinweg 55, 2333 CA Leiden, The Netherlands\label{aff164}
\and
Institute of Astronomy, University of Cambridge, Madingley Road, Cambridge CB3 0HA, UK\label{aff165}
\and
Space physics and astronomy research unit, University of Oulu, Pentti Kaiteran katu 1, FI-90014 Oulu, Finland\label{aff166}
\and
Center for Computational Astrophysics, Flatiron Institute, 162 5th Avenue, 10010, New York, NY, USA\label{aff167}}

\date{\today}

\authorrunning{Euclid Collaboration: W.~Gillard et al.}
\titlerunning{\ShortTitle}

\abstract
{
ESA's \Euclid cosmology mission relies on the very sensitive and accurately calibrated spectroscopy channel of the Near-Infrared Spectrometer and Photometer (NISP). With three operational grisms in two wavelength intervals, NISP provides diffraction-limited slitless spectroscopy over a field of 0.57\,deg$^2$. A blue grism \BGE\ covers the wavelength range 926--1366\,nm at a spectral resolution $\mathcal{R}=440$--900 for a 0\farcs5 diameter source with a dispersion of 1.24\,nm\,px$^{-1}$. Two red grisms \RGE\ span 1206 to 1892\,nm at $\mathcal{R}=550$--740 and a dispersion of 1.37\,nm\,px$^{-1}$. We describe the construction of the grisms as well as the ground testing of the flight model of the NISP instrument where these properties were established.
}

\maketitle

\section{Introduction}

\Euclid is a new space mission within the European Space Agency's (\acs{ESA}) {\it Cosmic Vision 2015--2025} programme, aiming to study the nature of dark matter and dark energy by estimating the expansion history of the Universe and the growth rate of cosmic structures \citep{laureijs2011,EuclidSkyOverview}. For this purpose \Euclid was designed as a wide-field telescope in the visible to \ac{NIR} wavelength range to survey $\sim$\,14\,000\,deg$^2$ of extragalactic sky from the Sun--Earth Lagrange point 2 \citep{Scaramella-EP1}.

To achieve its objectives, \Euclid's payload consists of two scientific instruments developed by the Euclid Consortium: VIS provides high-resolution images of the sky in a single broad wavelength band from 550\,nm to 950\,nm \citep{EuclidSkyVIS}. With a spatial sampling of $\ang{;;0.1}$ and a tight control on the \ac{PSF}, the VIS instrument is optimised for the measurement of galaxy shapes for weak lensing, which enables the mapping of dark matter structure as a tool for cosmological studies \citep{Cropper2016a,EuclidSkyVIS}.
\Euclid's second instrument, the \ac{NISP} \citep{EuclidSkyNISP}, operates in the \ac{NIR} and shares a common \ac{FoV} of 0.54\,deg$^2$ with VIS \citep{EuclidSkyOverview} thanks to a dichroic transparent to \ac{NIR} which is located in front of the NISP entrance pupil.
\ac{NISP} has been designed for the spectroscopic measurements of galaxy distances for the {galaxy clustering} probe, as well as \ac{NIR} information for photometric redshifts.

As a cosmology mission, \Euclid's ultimate goal is to reach a combined cosmological `\acl{FoM}' \acs{FoM}\,$\geq$\,400 \citep{EuclidSkyOverview} on the dark energy parameters $(w_\mathrm{p},w_a)$ with weak lensing and galaxy clustering. Reaching this ambitious goal requires a combined approach that leverages both a comprehensive joint analysis of weak gravitational lensing, galaxy-galaxy lensing, and galaxy clustering (commonly referred to as a $3\times2$-point analysis) and a three-dimensional galaxy clustering analysis using spectroscopic redshifts to evaluate cosmological distances.
To that end, the \Euclid spectroscopic survey is designed to target H$\alpha$ line-emitting galaxies in the redshift range 0.9--1.8, thereby providing precise measurements of the growth of structure and redshift-space distortions as well as the expansion history via baryon acoustic oscillations. Consequently, the galaxy clustering analyses requires to obtain spectroscopic redshift $z$ for at least 1700 galaxies per square degree.
This corresponds to redshift measurements for $\sim$\,2.55$\times$10$^7$ galaxies over the survey area of $\sim$14\,000\,deg$^2$ of the Euclid Wide Survey \citep{Scaramella-EP1} achieved through slitless spectroscopy.
The \Euclid core science also requires the galaxy redshifts to be estimated with an accuracy $\sigma_z\leq0.001(1+z)$, which translates to a minimal spectral resolution $\lambda/\Delta\lambda \geq 380$ when considering a source size of $\ang{;;0.5}$.

\begin{figure*}[!t]
\centering
    \includegraphics[width=0.65\textwidth]{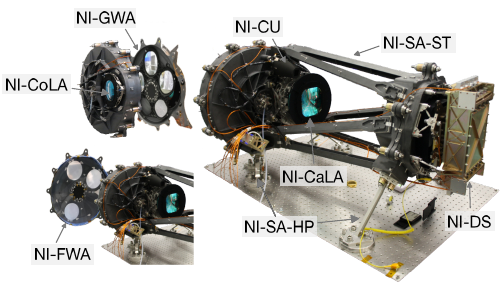}
    \caption{Schematic view of the NISP instrument detailing its sub-systems. {\it Right:} View of the whole instrument as integrated. {\it Left:} Detailed view with an open wheel cavity, showing the locations of the two wheels to select spectral elements. Not shown is the `warm' electronics for commanding and data processing that is located away from the main NISP instrument assembly in \Euclid's service module \citep{spie12.2232941}.The sub-system names and functions are described in the text.
    \label{fig:NISP}}
\end{figure*}

This work focuses on evaluating NISP spectroscopic performance through the analysis of \ac{PSF} quality and spectral dispersion properties.
In the following we will describe the NISP spectroscopic channel that was designed and built for \Euclid's \ac{NIR} spectroscopic survey. \Cref{sec:nispoverview} provides a general overview of the \ac{NISP} instrument as well as the grisms as the key components for NISP's spectroscopic capabilities. The setup for NISP ground tests of the spectroscopic channel are described in \cref{sec:testingground}. NISP's spectroscopic optical quality is presented in \cref{sc:IQ} and \cref{sec:dispersion} shows a detailed analysis of the grisms' dispersion properties. In \cref{sec:performance} the overall resulting spectroscopic performance is discussed.  The paper closes with a brief conclusion and a description of available data products characterising the grisms in \cref{sec:conclusions}.

\section{The NISP instrument}
\label{sec:nispoverview}

The NISP instrument \citep{10.1117/12.2630338,EuclidSkyNISP} is a near-infrared spectrometer and photometer whose spectroscopic channel has been designed and optimised to detect H$\alpha$ emission line galaxies in the redshift range $z \in [$0.9,\,1.8]. 

The NISP general layout and components are shown in \cref{fig:NISP} -- a full presentation of the instrument can be found in \citet{EuclidSkyNISP}, and the photometry channel and bandpasses are described in \citet{Schirmer-EP18}. Here we will provide a brief introduction into the central aspects relevant for spectroscopy.

\subsection{Overview of NISP components and properties}

NISP consists of a \ac{SiC} mechanical structure (\acl{NI-SA-ST} -- \acs{NI-SA-ST}) that houses the \ac{NI-DS} on one side and the optical system on the other \citep{bougoin2017}. The \ac{SiC} was chosen for its stiffness and its thermal stability allowing to maintain optical alignment and stability over the mission lifetime. Additionally, fine tuning of internal NISP heaters maintain the opto-mechanics temperature in the required range around \SI{135}{\kelvin}--\SI{140}{\kelvin} to further maintain optical alignment. Low conductance (\SI{0.043}{\watt\per\kelvin} in total) bipods and monopod (\acl{NI-SA-HP} -- \acs{NI-SA-HP}) interface the NISP instrument with the \Euclid \ac{PLM} baseplate.

The \ac{NI-DS} \citep{10.1117/12.2234658} acquires images by sampling the \ac{FoV} with 16 HgCdTe 2k$\,\times$\,2k pixels \ac{NIR} detectors produced by Teledyne Imaging Sensors, arranged in a 4\,$\times$\,4 mosaic.
Their pixel size of 18\,\micron\ provides NISP with a spatial sampling of the sky of $\ang{;;0.3}\,\text{px}^{-1}$, and the detectors have their wavelength cut-off tuned to 2.3\,\micron\ to minimise thermal noise from the optical system's thermal emission.
Maintaining the focal plane below \SI{95}{\kelvin} is crucial for optimal performance and is achieved through thermal coupling of the instrument with an external radiator using four highly conductive thermal straps. Furthermore, NISP's detectors are read out by 16 individual SIDECAR ASICs, each operating at approximately \SI{140}{\kelvin} and dissipating up to \SI{4.8}{\watt} of heat.
To manage this heat load effectively, high-efficiency thermal coupling to the baseplate is employed, facilitated by \ac{MHP}s. Due to the substantial heat dissipation, a large radiator of approximately \SI{2}{\square\meter} is required to radiate this heat into space, accounting for both the electronics' heat and any conductive or radiative leaks between the \ac{PLM} and the radiator.

The NISP instrument images the sky in two different channels: a photometric channel for the acquisition of images with broadband filters and a spectrometric channel for the acquisition of slitless dispersed images. The optical system of NISP \citep{10.1117/12.2529114,grupp2012} comprises several elements; 
a spherical-aspherical meniscus-type \ac{CoLA} at the entrance of NISP that actively takes part in correcting the chromatic aberration of the image; the \ac{NI-FWA} with a set of three filters \citep{Schirmer-EP18}, namely \YE\ (950--1212\,nm), \JE\ (1168--1567\,nm) and \HE\ (1522--2021\,nm) as well as a `closed' position, blocking light from the telescope;  the \ac{NI-GWA} with a set of three {red} grisms (\RGlow$-$\RGhigh\,nm), a single {blue} grism (\BGlow$-$\BGhigh\,nm) and an open position; 
and finally an assembly of three lenses which, together with their holding structure, constitute the \ac{CaLA}, focusing light onto the detector plane. 

\subsection{The NISP grisms}

The key components in creating dispersed light for determining redshifts of distant galaxies are the NISP grisms.
They are the combination of a transmission grating with a prism. This combination allows us to create a common \ac{FoV} with the NISP photometry channel, placing the dispersed light onto the same detector plane.
The NISP instrument contains four different grisms labelled RGS000, RGS180, RGS270, BGS000, three in a red and one in a blue bandpass. 

\begin{figure}[!t]
    \centering
        \includegraphics[width=\columnwidth]{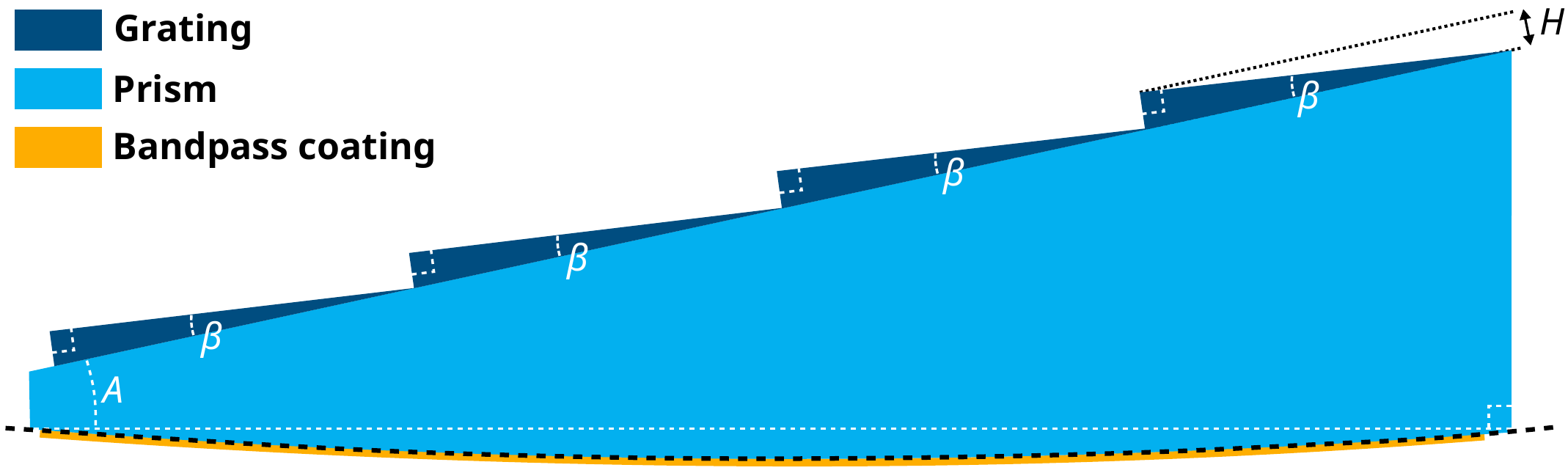}
    \caption{Cross-section of a NISP grisms. The prism wedge (light blue) has an added curved basis surface onto which the filter coating is deposited (yellow), while the grooves of the dispersion grating (dark blue) are directly engraved onto the flat hypotenuse surface of the prism. The grisms are characterised by their prism angle $A$, the groove height $H$, and the blaze angle $\beta$. \label{fig:grism-cutout}}
\end{figure}

Each of the grisms comprises several optical elements, each fulfilling a specific function \citep{costille2019}, presented in \cref{fig:grism-cutout}.
A resin-free blazed dispersion grating (dark blue) is directly engraved onto the hypotenuse surface of a fused-silica prism (light blue) to disperse the light. Each grism has a diameter of 140\,mm and a central thickness of approximately 12\,mm \citep{costille2014}, making them the largest grisms ever flown into space.   
The prism and the grating design were defined to maximise transmission efficiency at the undeflected wavelength of about 1.2\,\micron\ for the red and 0.9\,\micron\ for the blue grisms, with transmission required to be higher than $70\%$ in grism bandpass.
Additionally, the grating groove has been engraved with a curvature on the hypotenuse surface of the prism to apply a spectral wavefront correction to the incoming light.
The base of the prism carries optical power, that is, it has a curved surface (dashed line) to provide a fine-focus correction for the incoming light beam towards the NISP detection plane. 
Finally, a multilayer coating (yellow) has been deposited on the curved surface to define each grism's transmission bandpass. 

The grisms RGS000, RGS180 and RGS270 are three {red} grisms transmitting light in the spectral band \RGE, ranging from about \RGlow\,nm to \RGhigh\,nm, with these wavelengths marking half of the maximum of the in-band transmission.
The grism BGS000 is a {blue} grism with a spectral band ranging from about 880\,nm to \BGhigh\,nm. Combined with the transmission characteristics of \Euclid's dichroic and telescope mirrors, the \Euclid blue spectroscopic bandpass \BGE\ is reduced to \BGlow--\BGhigh\,nm. This work will focus on the grism characteristics themselves, as telescope mirrors and dichroic were not present during the instrument test campaigns. Their -- well known -- characteristics are simply folded in later.

\begin{figure*}[!t]
    \centering
    \includegraphics[width=0.9\textwidth]{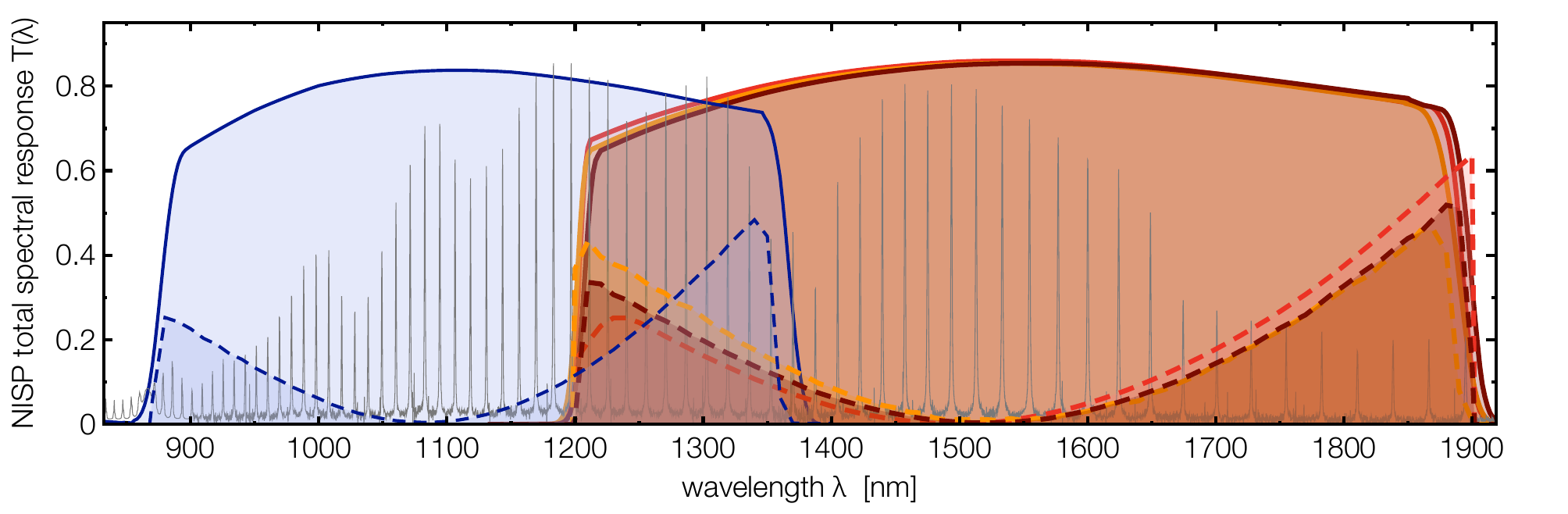}
    \caption{Total transmission of the spectra's $1^\text{st}$-order (solid line) and $10\times$ total transmission of the spectra's $0^\text{th}$-order (dashed line) for the BGS000 (blue), RGS000 (light red), RGS180 (orange), and RGS270 (dark red) grism channels. The total transmission accounts for the transmission of the grisms themselves, of \ac{CoLA} and \ac{CaLA}, and includes the detector quantum efficiency. For comparison, the Fabry--Pérot etalon spectrum (light grey) is shown as measured in the laboratory with a Lambda 900 spectrometer, in arbitrary data units. The contributions from neither the dichroic nor telescope are accounted for in these measurements as \Euclid mirrors as well as dichroic where not emulated during the instrument tests. \label{fig:Fabry-Perrot}}
\end{figure*}

The grisms create different orientations of spectral dispersion on the focal plane. With a view from the NISP optics towards the projected spectra on the detector plane and with grism RGS000 as the reference, the grism RGS180 has a 180$^\circ$ orientation and hence disperses light in the opposite direction to RGS000.
Similarly, RGS270 has a 270$^\circ$ counter-clockwise orientation with respect to RGS000. 
These three different orientations were chosen to permit `decontamination' of spectra from overlaps with other sources, creating a reconstructed spectrum from successively acquired exposures of a given field with the three red grisms.

Each grism is glued onto six Invar pads, whose coefficient of thermal expansion closely matches silica. The blades are bonded to flexible aluminium-alloy blades that limit stress transmission during the cooldown; these blades are part of the aluminium-alloy mount, which is directly integrated into the grism wheel \citep{SPIE844284421113}.

\section{NISP ground testing setup}
\label{sec:testingground}

After completion of the instrument integration in 2019, the NISP flight model underwent a series of tests in vacuum and with the instrument cooled down to its operational temperature. Two test campaigns were conducted inside the ERIOS vacuum chamber at the \ac{LAM} in the fall of 2019 and the beginning of 2020, to validate the instrument's performance and its functionality in a space environment. 

The  ERIOS chamber \citep{Costille2016b} is a large cryogenic chamber, with an external envelope of 6\,m in length and a diameter of 4\,m, capable of achieving high vacuum ($\sim$\,10$^{-6}$\,mbar) and to cool down an inner volume of 45\,m$^3$ to low temperatures ($\sim$\,80\,K) with a high stability ($\pm$4\,mK) thanks to liquid nitrogen shrouds covering the entire inner surface of the chamber.

In addition, various types of ground-support equipment were specifically developed by the NISP instrument team to enable the ground tests \citep{costille2017}.
One important component was a point-like light source developed to measure the NISP object plane and to verify NISP's optical performance, mainly plate scale across the detector plane, point-spread-function width, and a rough estimate of ghost images. The point-like source was the combination of a pin-hole light source and a telescope made of a 160\,mm diameter elliptical off-axis mirror with primary and secondary focal points at 500\,mm and 3000\,mm, respectively, simulating \Euclid's F/20 telescope beam.

The entire aperture of the mirror was illuminated through this 2\,\micron\ pin-hole located at the primary focus of the elliptical mirror, creating a point source for NISP.
Both, the mirror and the pin-hole were attached to the same baseplate and the whole system (mirror, pin-hole, and baseplate) was made from silica to ensure stability of the mirror's focal distance. 
The telescope baseplate was attached to translation and rotation stages using thermal flexures made from a low conductivity material, thermally isolating the telescope, while providing it with 5 degrees of freedom when being pointed towards different positions in NISP's \ac{FoV}. 

During the tests, this telescope was inside the ERIOS chamber, together with the instrument, and was operated at temperature of $\simeq160~\text{K}$ with stability of $0.04~\text{K}$ over a test duration of 1 month. The motors of the telescope mount were kept at warm ambient temperature and were isolated from the cold environment of the ERIOS chamber by a Multi-layer Insulation blanket. 

Inside ERIOS, light was fed to the telescope through a cryogenic fibre connected to the pin-hole, with light sources being located outside the chamber.
At the interface of the ERIOS chamber a pair of vacuum windows connected the fibre with the light sources, using a cooled neutral density filter to suppress thermal background. 
On the outside, a set of neutral density filters allowed us to select the intensity of light entering the fibre.

Different types of light sources were used during the instrument test campaigns.
Among them was a continuum laser source that was either connected to a monochromator, providing a selectable monochromatic wavelength in the range from 0.4\,\micron\ to 2.1\,\micron, or to a Fabry--Pérot etalon to emulate an emission-line spectrum with a total of 64 emission lines in the bandpass of the NISP grisms.  
Alternatively, an argon lamp could be used as a source to provide a well known reference spectrum for comparison with the Fabry--Pérot etalon.
The monochromator and the Fabry--Pérot etalon were both characterised before delivery to the ground test equipment.

Spectral lamps were fed into the monochromator to characterise and calibrate it, using HgCd, He, or Cs lamps for the visible band below \SI{1.1}{\micro\meter}, and an Ar lamp for the near-infrared domain. The output of the monochromator was scanned using either silicon or germanium photodiodes. The calibration yielded a selectable maximum wavelength accuracy of \SI{0.4}{\nano\meter}, and the monochromator bandwidth was measured to be approximately \SI{0.41}{\nano\meter} in \ac{FWHM}.

The etalon was re-aligned and re-calibrated in warm conditions with a Perkin Elmer Lambda 900 UV/VIS/NIR spectrometer (`Lambda 900') before testing of NISP started. During the tests, thermal sensor monitored the stability of the light sources and the etalon. 
\Cref{fig:Fabry-Perrot} shows the Fabry--Pérot etalon's spectrum as it was measured and calibrated with a $1\sigma$ precision of $\pm$0.2\,nm using the Lambda 900 spectrometer.
A comparison to the grisms's transmission shows the number of Fabry--Pérot's peaks available for calibration. The grism channels' total transmission shown here accounts for the transmission of the \ac{CoLA} and \ac{CaLA} lenses, the grisms themselves, as well as the detector quantum efficiency. In this figure neither the \Euclid mirrors nor dichroic are accounted for, as both elements were not present in the instrument-level test setup. Note that in \cref{fig:Fabry-Perrot}, the transmission of the $0^\text{th}$-order components of NISP's spectrograms, which are below few per cent at maximum, have been multiplied by a factor of 10 to make them stand out. Additionally, the optical transmission shown in \cref{fig:Fabry-Perrot} is derived from combined subsystem-level measurements. Due to limited knowledge of the optical ground equipment transmission -- including fibre, pinhole, and telescope contribution to the total transmission -- and the absence of a calibrated photodiode in the ERIOS chamber, we were unable to measure the absolute transmission of the NISP optics during the ground test campaign, and thus we relied on subsystem measurements to estimate the total transmission presented in \cref{fig:Fabry-Perrot}.

Prior to the start of the optical tests, the focal length of the NISP instrument was measured at cold (NISP focal plane at $\simeq85\,\text{K}$ and NISP optics at $\simeq135\,\text{K}$), using a monochromatic light beam with a wavelength of 1000\,nm in the \YE photometric passband.
In this process, \ac{PSF} widths were measured from spots created by the point source. This was carried out at five positions in the NISP \ac{FoV}, where hundreds of individual slightly dithered exposures were taken to over-sample the PSF and to increase statistical accuracy on the \ac{PSF} width measurement. This was repeated for different distances of the telescope simulator from NISP.
During the \ac{PSF}s acquisitions, the relative distance between NISP and the telescope simulator was accurately monitored with a laser tracker system that measured the position of ten targets attached to the NISP mechanical structure relative to five targets attached to the back of the telescope mirror \citep{costille2016,costille2018}. 

This set of measurements provided a reference focal plane position for the instrument in the \YE passband.
From these measurements an optimal object plane position was identified, using an as-built {\it Zemax} model\footnote{\href{ https://www.zemax.com}{www.zemax.com}} \citep{moore2004} of the NISP instrument, providing the best image quality in all NISP filter and grism bandpasses. 
All subsequent instrument tests where then made with the telescope simulator pin-hole positioned on this optimal object plane.

It is important to note that this measurement also established the fixed reference plane for positioning NISP relative to the VIS instrument, ensuring optimal NISP image quality once the Euclid telescope is focused on the VIS instrument, which has more stringent image quality requirements. The relative alignment between NISP and VIS was verified during the \ac{PLM} ground tests and is beyond the scope of this paper. Nevertheless, the excellent optical performance delivered by NISP in flight confirms the accuracy and stability of its alignment with the Euclid telescope.

\section{NISP spectroscopic optical quality\label{sc:IQ}}

\begin{figure*}[!t]
    \centering
    \includegraphics[scale=0.8]{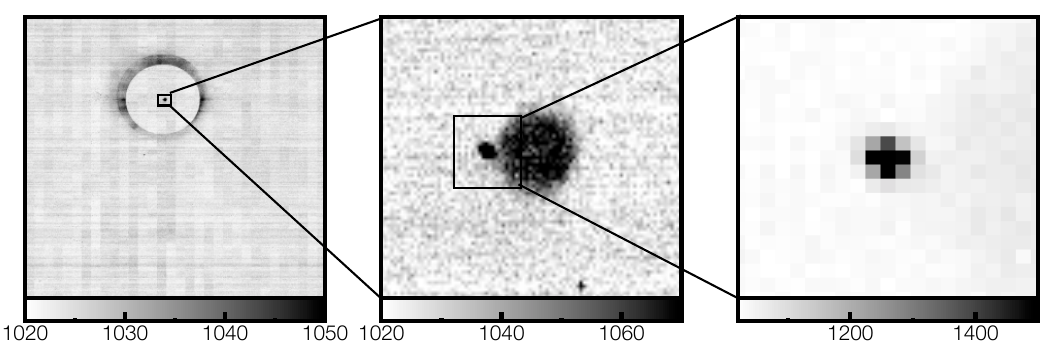}
    \caption{
    {\it Left}: 2k\,$\times$\,2k image of a single NISP's detector showing the $1^\text{st}$-order dispersed image of a monochromatic \ac{PSF} at 1800\,nm taken with NISP RGS000. The corresponding $0^\text{th}$-order isn't visible in this image because it falls onto detector gaps about 800 pixels above the $1^\text{st}$-order \ac{PSF}.
    {\it Centre}: Zoom-in showing the surrounding area of the \ac{PSF} in a region of 40\,$\times$\,40 pixels. {\it Right}: Zoom in on a region of 10\,$\times$\,10 pixels centred on the \ac{PSF}. Here, the colour scale has been adapted not to have the central core of the \ac{PSF} saturated as in the left and central image. For a detailed description of image features please see text.
    \label{fig:PSF_descriptor}}
\end{figure*}

This section describes the optical quality of the NISP spectroscopic channel, when a grism is moved into the NISP science beam. First,  RGS000, RGS180, and BGS000 are discussed, then the special case of RGS270, for which a non-conformance has been identified during instrument tests, rendering it in-operational for science use.

\subsection{Optical quality of RGS000, RGS180 \& BGS000\label{sc:NISP-S_IQ}}

The NISP optical quality was assessed by evaluating the radii that encircle 50\% and 80\% of the \ac{PSF}'s total energy -- in the following these radii will be referred to as EE50 and EE80, respectively.
In the spectroscopic channel, this verification was done for every grism of the \ac{NI-FWA} at three monochromatic wavelengths: 1300\,nm, 1500\,nm, and 1800\,nm for all three red grisms and 900\,nm, 1120\,nm, and 1340\,nm for the blue grism.
Additionally, we tested image quality of both RGS000 and RGS180 with a $\pm\ang{4}$ grism wheel rotation offset to validate that offsetting the grism positions by this amount would not degrade image quality. 
This additional measurement was part of an evaluation of a modified survey strategy to overcome the reduced quality of data from RGS270 (see \cref{sec:rgs270}).

The image quality was evaluated with an analysis of monochromatic \ac{PSF} images taken at the four corners of the NISP \ac{FoV}.
Because the NISP plate scale of $\ang{;;0.3}$ per pixel undersamples the \ac{PSF}, hundreds of \ac{PSF} images were acquired around each position, dithering by a tenth of a pixel around the target position in the NISP image plane. 
This dithering, obtained by moving the telescope on its axes, allowed to spatially over-sample the \acp{PSF} and reduced the statistical uncertainties on the EE50 and EE80 measurements. 

Before measuring \ac{PSF} properties, images acquired by the NISP instrument were corrected for pixel quantum efficiency and conversion gain, and a NISP residual {thermal} background acquired under `dark test conditions', i.e.\ with the NISP \ac{NI-FWA} in closed position and all light sources turned off.
An automatic procedure, derived from the work of \cite{10.1117/1.OE.55.6.063101}, was used to locate the \ac{PSF} within the 2k$\,\times$\,2k pixels of the target detectors.
Once located, a 20\,$\times$\,20 pixel stamp centred on the expected \ac{PSF} position was extracted from the images and \ac{PSF} metrics were extracted by image analyses.
Every extracted sub-image was controlled by eye and e.g.\ detector cosmetic defects misidentified as \ac{PSF}s by our algorithm were rejected from the analyses. 
An \ac{IPC} correction, relying on a Richardson--Lucy deconvolution method \citep{lucy1974,richardson1972} using \ac{IPC} coefficients previously measured during detector characterisation as the deconvolution kernel, was applied to each extracted stamp.
However, it should be noted that the spread of the measured EE50 and EE80 distributions were much wider than the correction factor applied through this \ac{IPC} deconvolution. The \ac{IPC} signal leaking to neighbouring pixels was measured to be very small, of 2.87$\pm$0.01\% losses in the central pixel that are non-equally redistributed to the eight immediately adjacent pixels \citep{legraet2022}. 

\Cref{fig:PSF_descriptor} shows one example of a single exposure of a NISP \ac{PSF} acquisition.
The left panel is a full frame image (i.e.\ 2k\,$\times$\,2k pixels) from one detector, and the artificial point source at the top. A ring-like structure is visible surrounding the \ac{PSF}, that resulted from scattered light at the exit of the pin-hole. Since the ring radius is very large compared to the \ac{PSF} diameter, it was easily disregarded in the analyses.
However, these rings were quite useful in validating that the blind search algorithm indeed successfully located the \ac{PSF}. 
In the centre panel of \cref{fig:PSF_descriptor} a zoom into the small box around the \ac{PSF} is shown. The diffuse circular feature to the right of the compact \ac{PSF} is produced by light emission coming from the fibre core passing through a non-perfectly blocking neutral density filter around the pin-hole.
This was meant to block light throughout the full NISP wavelength range. It however appeared that instead it had a transmission increasing with wavelength, starting with an optical density $>$\,5 at 900\,nm and reaching an optical density of $\simeq$\,4.5 at 1800\,nm, creating this patch.  The rightmost panel shows another zoom step into the \ac{PSF}, with contrast adjusted to prevent display saturation.

The EE50 and EE80 are deduced from \ac{PSF} model fitting with the following model

\begin{eqnarray}
    S(y,z)&=&\int_{y-\delta_y/2}^{y+\delta_y/2}\int_{z-\delta_z/2}^{z+\delta_z/2} G(y',z',\vec{\hat{\theta}})\,\mathrm{d}y'\mathrm{d}z' \notag \\
    &+& C(y,z,\vec{\hat{\Theta}})+B\,,\label{eq:psf}
\end{eqnarray}
where $(y,z)$ are the spatial coordinates in the pixel mosaic reference frame, $G(y,z,\vec{\hat{\theta}})$ is a \ac{PSF} model of parameters $\vec{\hat{\theta}}$ -- for the functions that were considered see below -- $C(y,z,\vec{\hat{\Theta}})$ is a $C\infty$-Bell function \citep{boyd2006} of parameters $\vec{\hat{\Theta}}$ -- which is used to model the contribution from the fibre core -- and $B$, a constant introduced to account for a residual constant background. As the fit is limited to a small stamp of 20\,$\times$\,20 pixels centred on the \ac{PSF}, a constant background is a good first approximation at these scales. Note that the $z$ axis is defined such that the light dispersed by RGS000 along a spectral trace has its wavelength increase with $z$. 

As shown in \cref{fig:PSF_descriptor}, the NISP optical design concentrates most of the photon energy within a single pixel, yielding a \ac{PSF} with a \ac{FWHM} on the order of $0.7~\text{pixel}$. Such a small \ac{PSF} makes the estimation of EE50 or EE80 particularly challenging. Thus to estimate the reliability of our estimates, we tested three different functions $G(y,z,\vec{\hat{\theta}})$ to model the \ac{PSF} profile.
\begin{itemize}
    \item An {\bf asymmetric-Gaussian} function for which amplitude, centroid position, width along the $y$ and $z$ axes as well as inclination with respect to the $y$ axis are free parameters. When using this model, EE50 is evaluated as $\mathrm{EE50}_\perp=\sigma_\perp\sqrt{2\ln(2)}$, with $\sigma_\perp$ the width of the \ac{PSF} in the cross-dispersion direction. We used the cross-dispersion direction to minimise the impact of the monochromator bandwidth on the estimate of EE50. However, such an estimation might be underestimating the true width, by assuming the \ac{PSF} is having a Gaussian profile and neglecting potential contributions from a non-Gaussian tail.
    \item The sum of two asymmetric-Gaussian functions, hereafter refereed as a {\bf dual-Gaussian} profile, for which the respective amplitude, common centroid position, individual width along the $y$ and $z$ axes, as well as common inclination are free parameters. To avoid any degeneracy in the model fitting, the amplitude and width of the {\it larger} Gaussian was defined by a multiplicative factor relative to that of the {\it narrower} Gaussian. This model tends to better account for potential non-Gaussian tails in the \ac{PSF} profile. 
    \item An elliptical {\bf Moffat} profile \citep{moffat1969,serre2010} for which amplitude, scale factor, ellipticity, and inclination with respect to the RGS000's cross-dispersion direction are free parameters. This model also include non-Gaussian tails in the measured \acp{PSF}.
\end{itemize}

\renewcommand{\topfraction}{1.}
\renewcommand{\bottomfraction}{0.}
\renewcommand{\floatpagefraction}{1.}

\begin{figure}[!b]
    \centering
        \includegraphics[width=\columnwidth]{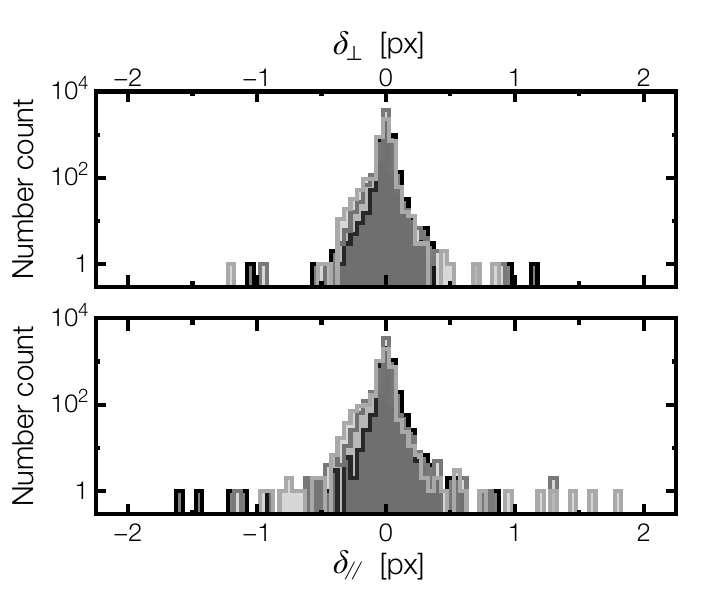}
        \caption{Differences in the measured position of the \ac{PSF} centroid for the various models used. The top panel shows the difference in position in the cross-dispersion direction $\delta_\perp$ while the bottom panel show the difference in position in the dispersion direction $\delta_\parallel$ -- in both cases the number count axis is logarithmic. The black histogram is comparing the elliptical Moffat profile to the dual-Gaussian profile. The grey histogram is comparing the asymmetric-Gaussian profile to the dual-Gaussian profile. The light-grey histogram is comparing the asymmetric-Gaussian profile with the elliptical Moffat profile.\label{fig:xy_diff}}
\end{figure}

Because the asymmetric-Gaussian profile may not be able to catch the faint extended tail of the \ac{PSF}, we did not use an analytical estimate of EE80, as we used for the EE50, but instead evaluated the EE80 after having subtracted the contributions from the fibre-core and background in the image, both derived from the model fitting. In this case EE80 is estimated by searching for the radius of the circle -- centred onto the \ac{PSF} centroid position -- that encapsulates 80\% of the total signal in the {\it background-free} image.
However, when working with the two other models, to avoid introducing errors when subtracting the background or fibre-core model from the \ac{PSF} image, the EE50 and EE80 were instead evaluated on the fitted model. 
In those cases, the radii were determined by evaluating the apertures containing 50\% and 80\% of the total signal from the \ac{PSF} model, using either the dual-Gaussian or Moffat components of the model, and excluding both background and fibre-core contributions.
However, as the monochromator did not have infinitely narrow bandwidth, and hence \acp{PSF} are slightly extended along the dispersion direction, both the EE50 and EE80 are slightly overestimated as this estimation disregards \ac{PSF} asymmetry.

The fits of the \ac{PSF} using the models presented in \cref{eq:psf} are performed on each individual \ac{PSF} exposure. The spatial sampling provided by dithering allows us to derive the statistical distribution of the \ac{PSF} model parameters, which will be discussed in the following paragraph.

The reduced $\chi^2$-distributions of the three models have a median value of $\simeq$\,1.04 for the asymmetric-Gaussian model, $\simeq$\,0.97 for the dual-Gaussian model, and $\simeq$\,1.02 for the Moffat profile. These values are too similar to prefer one model over the others. For every model a few cases show $\chi^2$ values larger than 1000. These were found to be correlated between the three models and are associated with the \ac{PSF} being located close to hot or pixels with `defects' like cosmic ray hits, Random Telegraph Noise, etc...\ that were not properly accounted for in our data reduction.  
We subsequently excluded these cases with biased fit results, removing \acp{PSF} with reduced $\chi^2 > 5$ which result in excluding about 5\% of the \acp{PSF}. Additionally, we rejected any \ac{PSF} with an unphysically large maximum signal $>$10 000\,e$^-$\,s -- given the photon flux provided by the optical ground equipment setup. 

\begin{figure}[t]
    \begin{subfigure}{\columnwidth}
        \centering
            \caption{Blue grism BGS000} \label{fig:fig:FoV_IQa}
            \includegraphics[width=0.76\columnwidth]{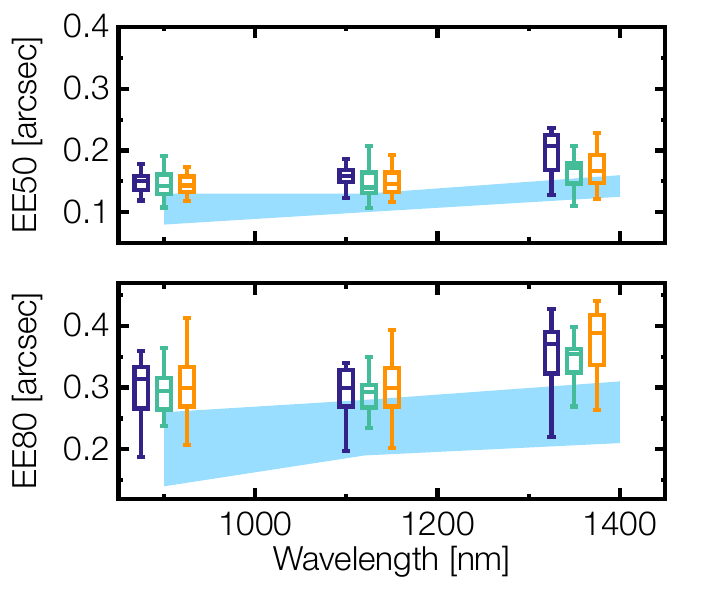}
        
    \end{subfigure}
    \begin{subfigure}{\columnwidth}
        \centering
            \caption{Red grism RGS000} \label{fig:fig:FoV_IQb}
            \includegraphics[width=0.76\columnwidth]{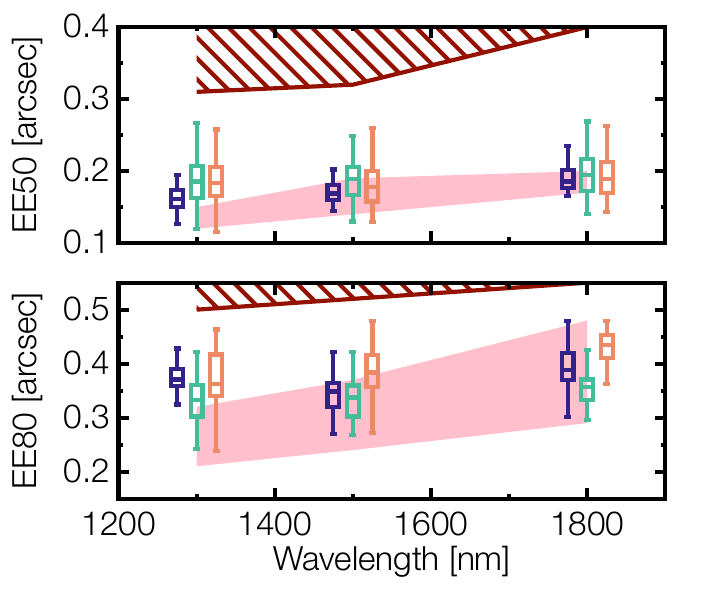}
        
    \end{subfigure}
    \begin{subfigure}{\columnwidth}
        \centering
            \caption{Red grism RGS180} \label{fig:fig:FoV_IQc}
             \includegraphics[width=0.76\columnwidth]{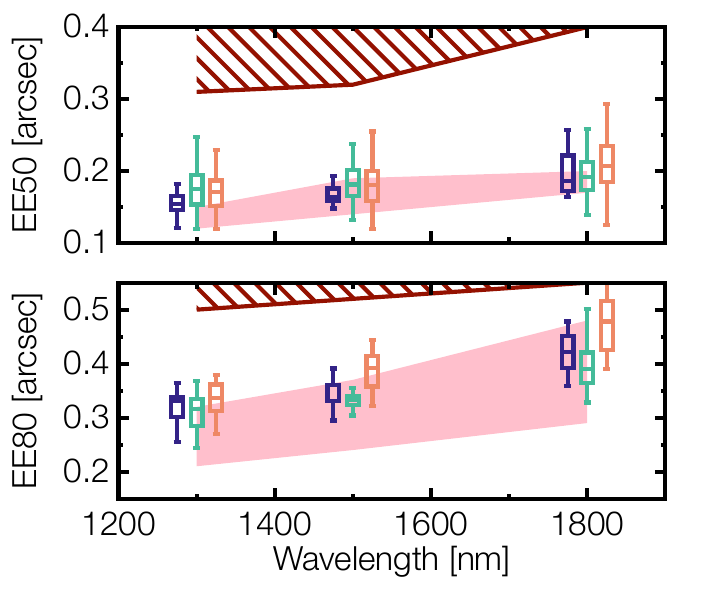}
        
    \end{subfigure}
    \caption{NISP grism image quality as function of wavelength, compared to the expectation from an ideal model limited by optical diffraction (shaded region) and to NISP scientific requirements (dashed region), for the grisms BGS000, RGS000, and RGS180. For each grism, the upper panel shows the EE50 estimated with the asymmetric-Gaussian (dark blue), dual-Gaussian (green), and Moffat profile (yellow). The bottom panel shows the corresponding EE80. An artificial wavelength shift was applied for plotting to increase readability.\label{fig:FoV_IQ}}
\end{figure}

\begin{figure}[!t]
    \centering
        \includegraphics[width=0.76\columnwidth]{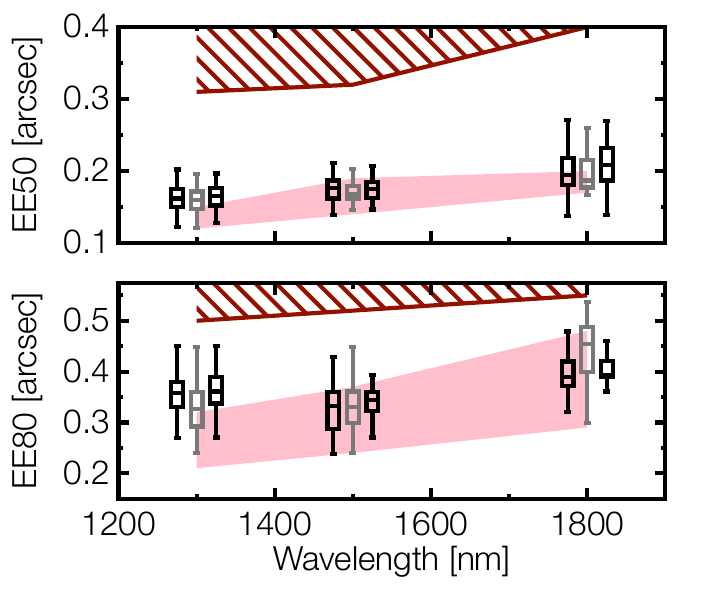}
        \caption{No image quality change between nominal and $\pm$\,\ang{4} offsets. Compared are EE50 (top panel) and EE80 (bottom panel) for nominal and offset positions, combining data for RGS000 and RGS180.  At 1300\,nm, 1500\,nm, and 1800\,nm the central grey boxes correspond to the \ac{EE} measured with the grism wheel in nominal position. The boxes to its left and right correspond to measurements done with $-\ang{4}$ and $+\ang{4}$ offsets, respectively. An artificial wavelength shift was applied for plotting to increase readability. For clarity the plot is restricted to the \ac{EE} estimate using the asymmetric-Gaussian profile.\label{fig:FoV_IQ_tilt}}
\end{figure}

\Cref{fig:xy_diff} shows the relative differences in the measured position of the \ac{PSF} between the different \ac{PSF} models.
With a standard deviation lower than 0.1 pixels and a mean of (7--10)\,$\times$\,10$^{-3}$ pixels, depending on the histogram, we concluded that all three models predict the same \ac{PSF} centroid position with an error on the order of one tenth of a pixel.
As those three models locate \ac{PSF} at roughly the same position, it is safe to consider they all identify the same \ac{PSF} which allows a comparison of the three profiles.

Despite the differences in the profile definition, the three models provide similar \ac{EE} values. Given the goal to assess the NISP spectroscopy image quality from \ac{PSF} size we conclude that these models overall provide robust estimates.
\begin{figure*}[!btp]
    \centering
    \includegraphics[width=0.95\textwidth]{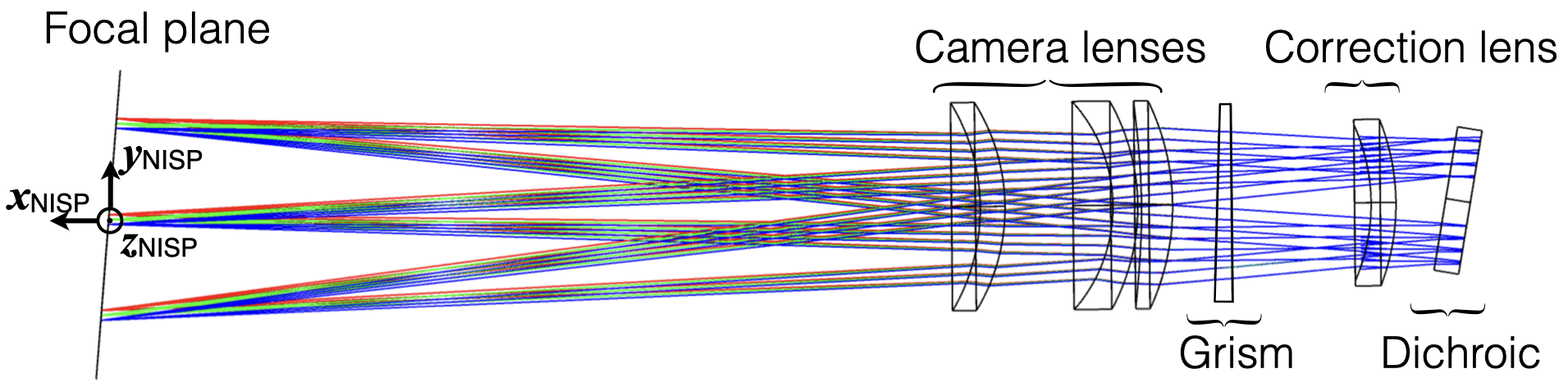}
    \caption{Ray tracing of the \ac{NI-OA} for three sources located at the centre and at two opposing edges of the focal plane. Light rays are shown, for each source, at wavelength of 1250\,nm (red), 1550\,nm (green) and 1850\,nm (blue), dispersed by grism RGS270. The dichroic element is located in the pupil plane of the telescope outside of NISP and is not part of the instrument.\label{fig:nisp_opt}}
\end{figure*}

\begin{figure*}[!t]
    \centering
    \parbox{0.9\textwidth}{\centering\includegraphics[width=0.9\textwidth]{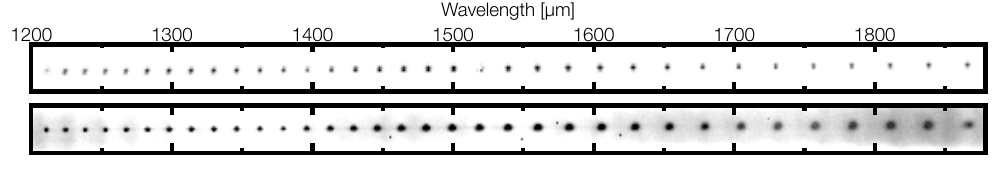}}\par
    \parbox{5.6cm}{\centering\includegraphics[scale=0.9]{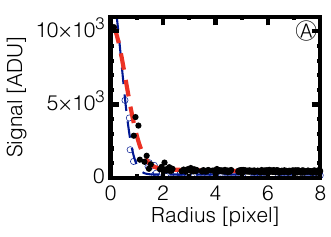}}
    \parbox{5.6cm}{\centering\includegraphics[scale=0.9]{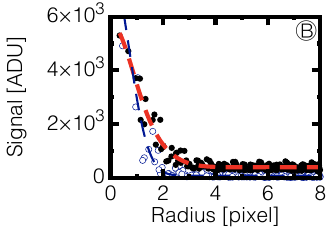}}
    \parbox{5.6cm}{\centering\includegraphics[scale=0.9]{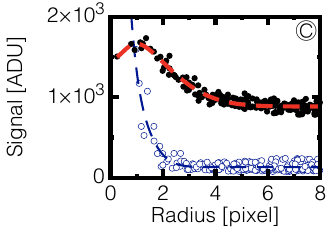}}
    
    \caption{{\it Top:} Image of the Fabry--Pérot etalon spectrum acquired with the NISP grism RGS000 during the instrument test campaign. {\it Middle:}~Image of the Fabry--Pérot etalon spectrum acquired with the NISP grism RGS270 during the instrument test campaign. {\it Bottom:} Spot size of the Fabry--Pérot emission lines displayed as radial profiles at 1300\,nm (A), 1540\,nm (B), and 1840\,nm (C). Measurement points are marked by open circle for RGS000 and dots for RGS270. The dashed lines are radial-profile fit with a double Gaussian to emphasise the shape of the \ac{PSF}. This functional form allows one of the two Gaussian to have a negative amplitude needed to reproduce the ring-shape of the RGS270's \ac{PSF} at large wavelengths.\label{fig:rgs270_spec}}
\end{figure*}

In \cref{fig:FoV_IQ} the overall image quality is shown, expressed as mean and spread of \ac{PSF} EE50 and EE80 across the NISP \ac{FoV}, derived from the three \ac{PSF} models. These are compared to an ideal optical diffraction-limited system (shaded area) and to the scientific requirements (dashed area). The boxes for each model represent mean, and first and third quartile of their values across the \ac{FoV} -- they are the combined result of the accuracy of our measurement and the actual \ac{PSF}-size variations. Similarly, the width of the ideal model band shows the variations of the theoretical \ac{PSF} across the \ac{FoV}.
One can see from this figure that the image quality of the NISP grisms is close to the theoretical expectation from an ideal system, with EE50\,$\simeq$\,0.6 times smaller than the scientific requirements $<$\ang{;;0.3} at 1500\,nm. A more detailed investigation of the four target fields shows variations in the median value of both EE50 and EE80, suggesting a smooth variation of \ac{PSF} size across the \ac{FoV}. However, these are smaller than the spread due to measurement uncertainties and are being disregarded.

To make sure that NISP grisms could also be operated with a rotational positioning offset of $\pm$\,\ang{4} without an impact on image quality (for the rationale see next section), additional acquisition with tilted grisms were made during the ground test campaign. \Cref{fig:FoV_IQ_tilt} compares the image quality for RGS000 and RGS180 for their nominal position and $\pm$\,\ang{4} offset. There are no differences. For clarity, this figure is only using the asymmetric-Gaussian profile as it has the smallest dispersion in the estimated EE50 -- but since all models lead to similar estimates of EE50, this result also holds for the other models.

One has to keep in mind that the data presented here were taken without the \Euclid telescope system, and hence the telescope \ac{PSF} has to be accounted for when extrapolating the measured NISP performance to in-flight conditions. Ground test of the \Euclid payload module, which contains both instruments as well as the whole \Euclid optical system, were conducted in summer 2022 at \ac{CSL} by Airbus Defence and Space. Although these tests are outside the scope of this paper, we can report that while \Euclid's telescope was prone to gravity which stressed the mechanical structure of the primary mirror, analysis did not reveal any degradation beyond expectations.

\subsection{Addressing RGS270 Non-Conformity}
\label{sec:rgs270}

One consequence of the \Euclid telescope \citep{refregier2010,laureijs2011,EuclidSkyOverview} having an off-axis three-mirror Korsch design \citep{korsch1977} is that in order to obtain high-quality images at all positions the NISP focal plane has to be tilted by an angle of \ang{4.8267} with respect to the optical axis \citep{EuclidSkyNISP}, as illustrated in \cref{fig:nisp_opt}.
The grisms RGS000 and RGS180 are dispersing light in the direction parallel to the $\vec{z}_{\text{NISP}}$ axis, i.e.\ perpendicular to the focal plane tilt. For this reason the focal length for exactly (and only) these two specific dispersions directly is identical for all wavelengths of a given object, as in an on-axis telescope.
On the other hand for grism RGS270, with a dispersion direction perpendicular to RSG000 and RSG180, the dispersion is running in a direction perpendicular to the $\vec{z}_{\text{NISP}}$ axis, i.e.\ in the direction of the focal plane tilt.
The curvature of the grating grooves in the RGS270 grism design was introduced specifically to correct for the tilt of the NISP focal plane, creating in-focus images on the focal plane at any wavelength.  

However, during our ground test campaign at instrument level we observed that images from grism RGS270 were partially defocused. 
This can be seen in \cref{fig:rgs270_spec} which compares Fabry--Pérot spectrograms acquired with RGS000 and RGS270 during NISP ground test campaign.
Focusing on the middle image presenting the RGS270 2D spectrogram, one can see that the bluest part of the spectrum (left-hand side) has a narrow and well focused \ac{PSF}.
In contrast to this, with increasing wavelength towards the right, the \ac{PSF} gets successively wider and at the reddest end even shows a ring-like appearance. There is clearly an increasing defocus with increasing wavelength.

After investigations, we concluded that the RGS270 grism was built using an improper interpretation of its optical design: the definition of the RGS270 optical reference frame was not properly propagated to the NISP mechanical reference frame -- which are oriented {\it inversely} to each other.
The main consequence of this error resulted in all models built for RGS270 to disperse light in the opposite direction to the design definition. As a result the focus compensation by the curved grating groove was going in the wrong direction.

A `Tiger Team', composed of Euclid Consortium members, investigated potential hardware solutions, including various possibilities to replace the RGS270 grisms that were built, as well as options to modify \Euclid's surveys.
After an assessment of both the risks for the mission induced by dismantling the NISP instrument as well as the time required to build a new RGS270 grism, a {\it survey solution} was proposed by the consortium and approved by the \acs{ESA}. This was based on a detailed analysis of a modification of survey and observation parameters, which we will discuss in the following.

To obtain the additional spectral orientation angles needed for decontamination of overlapping spectra, the survey solution involved to modify the survey strategy using NISP settings that rotate the \ac{NI-GWA} by $\pm$\,\ang{4} away from its nominal positioning of the grisms with respect to the centre of the beam, when observing either with RGS000 or RGS180. This in turn creates spectra tilted by $\pm$\,\ang{4} vs.\ the nominal orientation of spectra from RGS000 and RGS180.
After an in-depth simulation of different survey strategies we ended up with an optimal spectroscopic observation sequence. Each field is observed four times, with grisms used in the following order: RGS000\,$\rightarrow$\,RGS180+\ang{4}\,$\rightarrow $\,RGS000--\ang{4}\,$\rightarrow$\,RGS180. 

The advantage of this observation sequence, executed in the `K-pattern' of telescope dithering offsets for each sky position \citep{Scaramella-EP1,EuclidSkyOverview}, creates a geometry formed by the stacked spectra that offers even more dispersion angles than the initial observation sequence, for disentangling spectra during the offline data processing on ground. 
The largest drawback is that by slight rotation angle of the grism wheel, the nominal aperture of the grism -- defined by a baffle installed in front of the grating -- is shifted away from the nominal beam centre, introducing slight vignetting at the edge of the \ac{FoV}. 
This vignetting has been estimated from Zemax simulation to result in 10\% flux losses at \mbox{$\pm$\ang{0.385;;}} field angle, i.e.\ distance from the centre of the \ac{FoV}, which is reduced to $0\%$ flux losses at field angle \mbox{$<\ang{0.17;;}$}. 
On-sky simulation, involving the official \Euclid data reduction pipeline, demonstrated that this vignetting level is largely compensated by the detector quantum efficiency and an optical throughput which are both higher than scientifically required, and that the spatial structure can also easily be accounted for in the pipeline.
Larger tilts were also investigated but the increasing vignetting level quickly reduces the options for modifying the \Euclid survey strategy along this line.

As a conclusion, the RGS270 is kept on board the NISP instrument to maintain the centre of gravity during mission operation, but it will not be used for \Euclid's surveys. While an extensive dataset was acquired for RGS270 during ground testing, in the following the performance of RGS270 will not be discussed any further.

\section{NISP spectroscopic dispersion} 
\label{sec:dispersion}

An important part of the NISP ground test campaign was to establish a preliminary assessment of the grism's dispersion law as mounted inside the instrument, in preparation for the instrument calibration in-flight and NISP's in-flight operations. This section is detailing the procedure used to calibrate the NISP spectral dispersion on ground, explaining how the calibration datasets were acquired and analysed, before we discuss the validity and precision of this calibration.

\subsection{On-ground spectral dispersion calibration\label{sc:FP-cal}} 

To estimate spectral dispersion functions of the NISP grisms, the Fabry--Pérot etalon spectrum was projected with the telescope simulator to 144 individual points on the NISP \ac{FoV}. For every pointing a short exposure of $\simeq$\,13\,s was taken before the telescope simulator moved to the next field point. Additionally, at all positions corresponding to the centre of a NISP detector, we also used the Argon spectral lamp to obtain spectra as reference and control.

To avoid any bias induced by a small but still different alignment of the grism wheel after multiple repositioning, all measurements at these 144 points were made without turning the grism wheel. Only once a 144-pointing scan with a given grism was completed, the grism wheel was moved to position the next grism. This test sequence was used for all four grisms -- meaning the grisms BGS000, RGS270, RGS000, and RGS180 each at nominal positioning as well with the grisms RGS000 and RGS180 with a $\pm$\,\ang{4} rotation offset.  Also, from the point of view of the calibration, we chose to consider the rotated grism positions RGS000$\pm\ang{4}$ and RGS180$\pm\ang{4}$ to be independent grisms of their own and not to consider them to be identical to RGS000 or RGS180.

\begin{figure}[!t]
    \centering
        \includegraphics[width=\columnwidth]{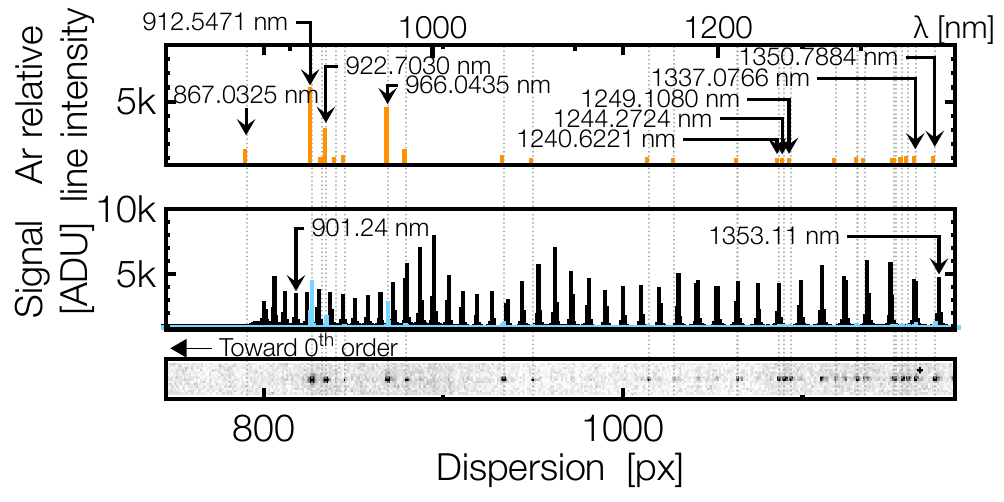}
        \includegraphics[width=\columnwidth]{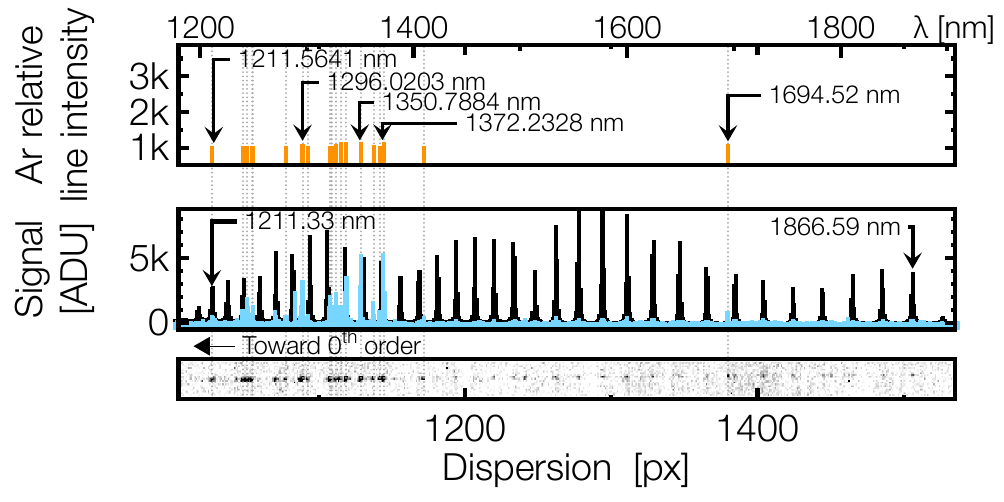}
        \caption{Calibration spectra for the NISP BGS000 (upper panel) and RGS000 grisms (lower panel). For each grism the 2D-image of the Argon spectral lamp spectrum is shown at the bottom, in the centre the Argon 1d-spectrum (cyan) compared to the Fabry--Pérot etalon spectrum (black), and at the top the reference Argon spectrum in vacuum from \cite{kramida2022}.\label{fig:BGSspectra}}
\end{figure}

Before going into the details of the calibration, \cref{fig:BGSspectra} presents examples of the $1^\text{st}$-order spectrogram for the blue and red grisms acquired during the NISP ground tests. The middle panels each show and compare spectrograms of the Fabry--Pérot etalon with those of the Argon spectral lamp. The top panels show reference Argon spectra in vacuum in these wavelength ranges, taken from the NIST database \citep{kramida2022}. By comparing the reference Argon spectra with those observed with NISP (middle and bottom panels) one can identify the strongest Argon emission lines present in both the \BGE\ and \RGE\ bandpasses. The middle panels also identify the Fabry--Pérot transmission peaks. For clarity, we limit the labelling to two peaks but based on the Fabry--Pérot calibration data all other transmission peaks were clearly identified\footnote{For reference see the calibrated Fabry--Pérot \acl{SED} in \cref{fig:grism-cutout}}. The middle and bottom panels do not cover the spectrogram of the  $0^\text{th}$-order which is located about 500 pixel away off the left-hand side. 

One note that, upon close inspection of the bottom panels, that both show a slight periodic excess. This is a charge persistence signal from the previous Fabry--Pérot etalon exposure. However, persistence did not affect the exposure of the Fabry--Pérot etalon itself, since the pointing position was changed between subsequent exposures and the Fabry--Pérot light source was brighter than the spectral lamp.

One central calibration for the NISP grisms is to describe the relation to convert wavelength for a source at any position into a position onto the focal plane. This is first established independently for each of the 144 spectrograms by modelling the 2d spectral trace with the following parametric function:
\begin{equation}
    \begin{array}{r@{\,=\,}l}
        \displaystyle y - y_0  & \displaystyle \sum_{i=0}^{m-1} C^y_i\,T_i(\lambda')\,, \\
        \displaystyle z - z_0  & \displaystyle \sum_{j=0}^{n-1} C^z_j\,T_j(\lambda')\,, \\
        \displaystyle \lambda' & \displaystyle \frac{\lambda - 0.5\left(\lambda_\text{max}+\lambda_\text{min}\right)}{0.5\left(\lambda_\text{max}-\lambda_\text{min}\right)}\,.
    \end{array}
    \label{eq:spectra}
\end{equation}
Here $(y,z)$ are spatial coordinates of a spectral feature on the focal plane, with $y$ and $z$ axis defined to be parallel to RGS000's cross-dispersion and dispersion direction, respectively. $(y_0,z_0)$ are spatial reference coordinates, $T_{i/j}(\lambda')$, are Chebyshev polynomials of the first kind defined in the $[\lambda_\text{min},\lambda_\text{max}]$ wavelength range, and $C^\kappa$ the Chebyshev coefficients, with $\kappa$ being either $y$ or $z$.

The $(y,z)$ coordinates of the Fabry--Pérot transmission peaks were measured by fitting the observed peaks with the \ac{PSF} profile described in \cref{sc:IQ}, providing position with an accuracy of $\simeq1/10^\text{th}$ of a pixel. 

This work uses as a reference the centroid position of the spectral $0^\text{th}$-order which is defined to be the position of the in-band $0^\text{th}$-order's wavelength with minimal transmission. This reference position is estimated from a template fit of the 2D image of the $0^\text{th}$-order spectrogram with a Fabry--Pérot template.

The transmission of the optical ground equipment has some residual knowledge uncertainties. To limit the impact of this, the Fabry--Pérot template is constructed by taking the averaged profile of its $1^\text{st}$-order spectrogram measured with the corresponding grism, corrected by the $1^\text{st}$-order transmission efficiency. In this way the template is considered to be representative of the photon flux at the entrance pupil of the NISP instrument. To account for the $0^\text{th}$-order, the template is then multiplied by the $0^\text{th}$-order transmission efficiency (see \cref{fig:Fabry-Perrot}) and convolved with a \ac{PSF} profile before being adjusted to the 2D images of the $0^\text{th}$-order. Parameters of the fit are the centroid position of the template, defined to be the central wavelength of the grism band-pass, where the $0^\text{th}$-order transmission efficiency is at its minimum, a stretch factor converting wavelength to pixel, the dispersion direction angle, and parameters of the \ac{PSF} profile defined in \cref{eq:psf} which are accounting for the contribution of the fibre core.

\begin{figure}[!t]
    \centering
        \includegraphics[width=\columnwidth]{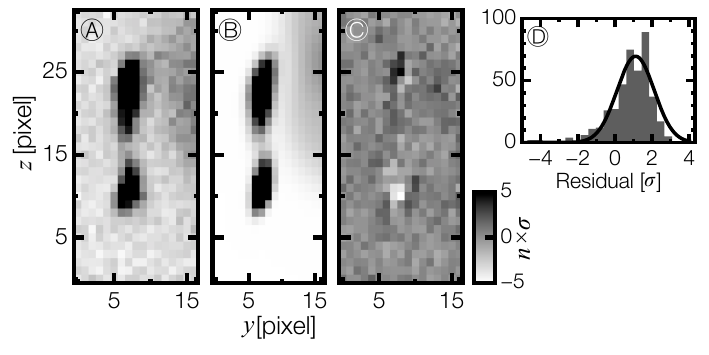}
        \caption{Example of the outcome of a template fit of the $0^\text{th}$-order spectrogram. Image in (A) is the image of the Fabry--Pérot $0^\text{th}$-order spectrogram as acquired by the RGS000 while (B) is the modelled image resulting from the template fit. Image (C) shows the residual of the template fit while (D) presents the residual distribution. \label{fig:zeroth_tmpFit}}
\end{figure} 

An example of such a template fit is shown in \cref{fig:zeroth_tmpFit} which presents the measured image of the Fabry--Pérot $0^\text{th}$-order spectrogram in (A) with the model in (B) and the residual in (C) and (D). Due to the prisms onto the NISP's grisms, the $0^\text{th}$-order are slightly dispersed, extending over several tens of pixels. Additionally, the grating's blaze function is optimised to maximise transmission in the $1^\text{st}$-order -- around \SI{1500}{\nano\meter} for the red grisms and \SI{1100}{\nano\meter} for the blue grism -- while minimising transmission at the same wavelengths for the  $0^\text{th}$-order. As a result, instead of appearing as a point-like feature, as would be expected for a purely dispersive grating, NISP's  $0^\text{th}$-order takes the form of an elongated, double-peaked structure. We tested the different \ac{PSF} profiles we listed in \cref{sc:IQ} and concluded that they all provide similar results without any of them outperforming the others. As for modelling the monochromatic \ac{PSF}, the usage of either of the profiles described in \cref{sc:IQ} led to similar centroiding of the $0^\text{th}$-order with negligible impact on the derived calibration parameters. The $0^\text{th}$-order position errors at $1\sigma$ were determined using $\chi^2$-profiling and subsequently propagated to the centroid position of the Fabry--Pérot transmission peaks during the calibration of individual $1^\text{st}$-order spectrograms. Analyzing the distribution of measured uncertainties in the $0^\text{th}$-order positions, we obtain mean uncertainties of $0.051\pm0.006,\text{pixels}$ with a standard deviation of $0.15$ pixels in the cross-dispersion direction and $0.12\pm0.02,\text{pixels}$ with a standard deviation of $0.4$ pixels in the dispersion direction. This highlights that positioning is more precise in the cross-dispersion direction, as the profile of the $0^\text{th}$-order is significantly thinner in this direction, leading to better constraint.

Coefficients of the Chebyshev polynomial as described in \cref{eq:spectra} are evaluated from a recursive $\chi^2$-fit. In the recursion, outliers were rejected at every iteration, each identified as spectral features located $>$\,5$\sigma$ away from the fitted spectral trace. Recursion stopped when no further outlier were identified. This allowed us to automatically reject some of the Fabry--Pérot transmission peaks that were insufficiently characterized by our \ac{PSF} modelling due to nearby hot or bad pixels.
Chebyshev polynomials were expanded up to the third order in the cross-dispersion direction and up to the fourth order in the dispersion direction. For the blue grism calibration the expansion was limited to the third order in both directions. The level of the expansion order was chosen in order for the averaged reduced $\chi^2$ to be the closest to $1$. With the above expansion order, the averaged reduced $\chi^2$ is of the order of $0.92\pm0.03$. However, we noticed that the $\chi^2$ distribution is skewed towards lower values. This is due to the limited accuracy of the template fit on a few of the $0^\text{th}$-order images which propagated to the $1^\text{st}$-order spectra and leads to large uncertainties in the relative position errors of some peaks' positions, $(y-y_0)$.

Up to this point, the Fabry--Pérot spectrograms were modelled independently of each other, leading to calibration coefficients $C_i^{\kappa}$ that smoothly vary across the \ac{FoV} depending on the location of the $0^\text{th}$ order. To complete the calibration, the spatial dependency of each of the $C_i^{\kappa}$ coefficients is then fitted by a bi-dimensional Chebyshev polynomial defined as 

\begin{equation}
        C_i^{\kappa}(y_0,z_0)=\sum_{i=0}^{m-1}\sum_{j=0}^{n-1}a_{ij}T_i(z')T_j(y')
        \label{eq:2Dcal}\,,
\end{equation}
with $a_{ij}$ the calibration parameters to be evaluated and $T_{i/j}(\kappa')$ the Chebyshev polynomials of the first kind, with $\kappa'$ being either $y'$ or $z'$ defined in the $[\kappa'_\text{min},\kappa'_\text{max}]$ range as

\begin{equation}
    \kappa' = \frac{\kappa - 0.5\left(\kappa_\text{max}+\kappa_\text{min}\right)}{0.5\left(\kappa_\text{max}-\kappa_\text{min}\right)}\,.
\end{equation}

Here $[\kappa'_\text{min},\kappa'_\text{max}]$ are the limits of the focal plane array, set to $[-85,+85]\,\text{mm}$ for both the $y$ and $z$ axes.
Again, the evaluation of the $a_{ij}$ parameters is carried out by a $\chi^2$ fit. The tables that summarise the calibration coefficients for each of the grisms are presented in \cref{app:calib}.
The advantage of \cref{eq:spectra} is that, through the set of coefficients $(C^y_i;C^z_j)$, it simultaneously describes the spectral dispersion and the curvature of the spectra, including non-linear effects induced by optical field distortion.
Furthermore, by modelling each $(C^y_i;C^z_j)$ coefficient over the entire field of view using \cref{eq:2Dcal}, one directly obtains a two-dimensional mapping of both dispersion and curvature through the higher-order coefficients $i$ or $j$ greater than or equal to 2.

\begin{figure}[!t]
    \centering
        \includegraphics[width=0.9\columnwidth]{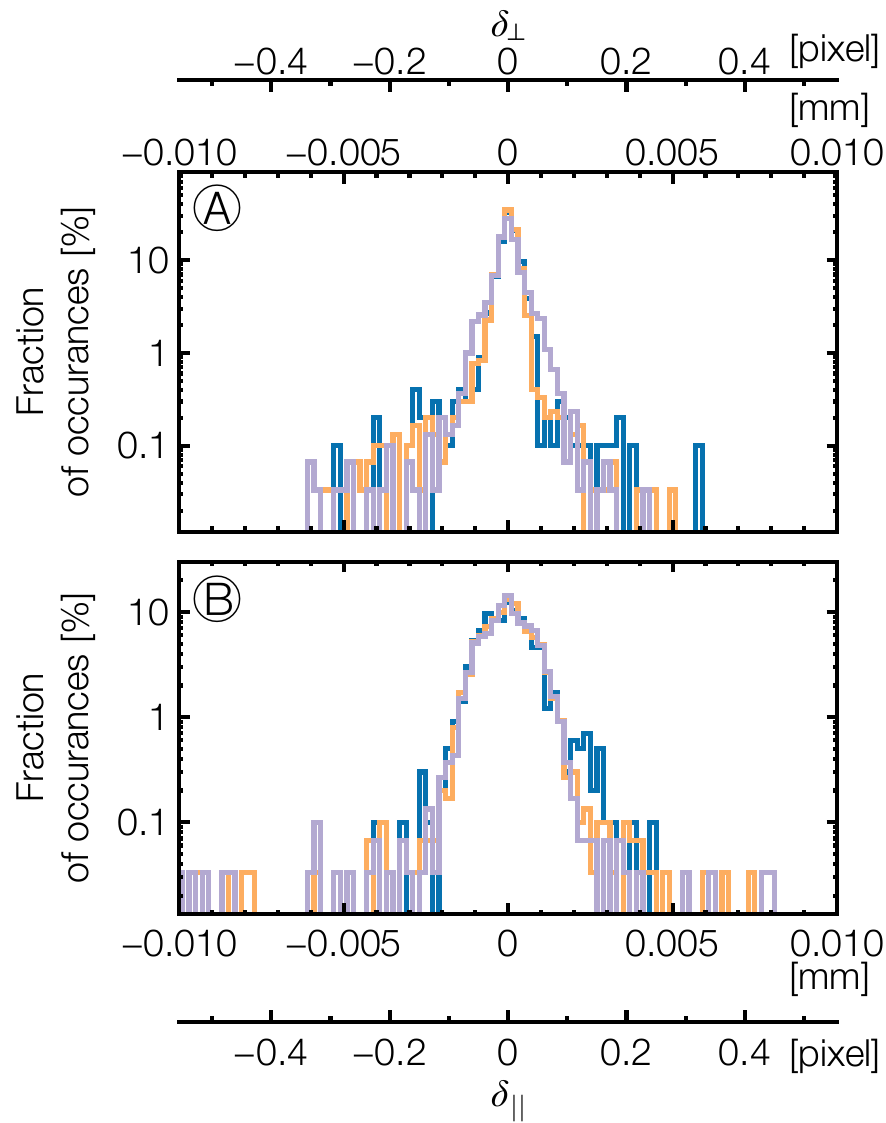}
        \caption{Normalised histograms of the distance between measured Fabry--Pérot emission lines and their positions predicted from calibration parameters in the cross-dispersion direction $\delta_\perp$ (A) and in dispersion direction $\delta_\parallel$ (B) for BGS000 (blue), combined RGS000 \& RGS000$\pm$4 (yellow) and combined RGS180 \& RGS180$\pm$4 (purple).}
        \label{fig:res_calib}
\end{figure} 

Theoretically, in a perfect optical system with constant grating groove, one expects the spectrograms to rotate around their undeflected wavelength when the grating is rotated around its centre. However, with the complexity of the NISP instrument itself, the mechanical uncertainties on the wheel position after its rotation, and on the pointing of the telescope simulators, it was not possible to precisely measure and confirm the position of the rotation centre for each of the spectrograms.
The choice was made to minimise the bias induced by the uncertainty on the location of each spectrum's centre of rotation by considering each grism's orientation as an independent grism to be calibrated. We tested the hypothesis that the rotation centre of the spectrograms was either their undeflected wavelength or the $0^\text{th}$-order\footnote{Which, strictly speaking, is not physically possible given the grism design, but this was nevertheless also tested, since at first approximation the $0^\text{th}$-order could be a proxy for the rotation centre.}. 
However, the calibration error derived from such a nominal position was found to be outside of our scientific requirement once it was applied to the rotated grisms, despite being calculated at sub-pixel level. 

Once we obtained the calibration parameters, we computed a predicted position of the Fabry--Pérot emission line using \cref{eq:2Dcal,eq:spectra} and compared the reconstructed and measured positions. This comparison is shown in \cref{fig:res_calib}. For clarity, we only show the results of the comparison for the blue grism BGS000 (in blue) and for the combined distribution of RGS000 and RGS000$\pm$4 (in yellow) as well as for the combined distribution of RGS180 and RGS180$\pm$4 (in purple). The calibration error we derived from this comparison is found to be lower than $7\times10^{-4}~\text{mm}$ (i.e. $<0.04$ pixel) in the cross-dispersion direction and lower than $1\times10^{-3}~\text{mm}$ (i.e. $<0.06$ pixel) in the dispersion direction for each grism. 

Additionally, we noticed that every spectrum for which the $0^\text{th}$ and $1^\text{st}$ orders were located on two different detectors shows a local offset in the estimated values of both the $C_y^0$ and $C_z^0$ coefficients, while the higher order coefficients of the parametrisation remained compatible with the coefficients estimated for any spectrogram fully falling onto a single detector. This offset behaves as if those spectra have a longer separation between the position of their $0^\text{th}$ and $1^\text{st}$ orders, breaking the otherwise rather smooth variation of the $C_y^0$ and $C_z^0$ coefficients across the field of view. Instead, we concluded that this offset has its origin in a bias induced by a limited precision of the detector metrology: this provides physical positions of the NISP detectors within the focal plane and was used to convert positions measured in pixels on a given detector to positions in the NISP focal plane reference frame. However, the detector metrology was only determined at room temperature by measuring  the position of reference marks engraved on the mechanical structure of each detector, using an optical camera mounted on a metrology bench. This method only provided a precision of the order of one pixel (i.e.\ $\simeq18\,\si{\mu\meter}$). The NISP focal plane metrology at operational temperature was therefore computed relying on the thermal model of the NISP instrument to predict, with limited precision, how thermal stress would modify the focal plane.

We try to re-calibrate the detector metrology at cold using the spectroscopic data themselves by aligning $1^\text{st}$-order spectra dispersed over multiple detectors. However, the small number of sources available (i.e.\ one single spectrum per NISP exposure in the entire NISP focal plane) did not allow a re-computation with sufficient precision for the spectroscopic calibration. We therefore left our cold reference-metrology as is and checked that the calibration parameters remain within 3$\sigma$ when computing calibration parameters with or without the spectra showing an offset in the $C_y^0$ and $C_z^0$ coefficients.
  
\subsection{Validation of the spectroscopic dispersion law} 

During the NISP ground test campaign, we acquired 16 spectrograms of the Argon spectral lamp with each of the BGS000, RGS000, and RGS180. An Argon spectrogram was observed every time the telescope simulator pointed to the centre of one of the 16 NISP detectors. This was achieved by replacing the Fabry--Pérot's fibre feeds within our optical ground equipment with feeds connected to the Argon spectral lamp. No other modifications were made to the instrument configuration or telescope simulator pointing. These Argon spectrograms were used to validate the spectroscopic dispersion law that we derived from the Fabry--Pérot etalon calibration.

Similarly to what was done for the Fabry--Pérot's spectrogram, we extracted Argon emission line positions based on model fitting of the 2D emission line profile. However, the detection process was significantly impacted by a substantial level of detector persistence charge signal from the previous Fabry--Pérot's exposure acquired at the same pointing as for the spectral lamp acquisition. A visual inspection of the initially identified line was conducted, to reject persistence signal peaks that are miss-identified as Argon emission lines by the automated peak-search algorithm, as well as ambiguous line detection. In the cases where it was challenging to differentiate between actual emission lines and persistence signal, the line was masked from the analysis.

Additionally, we extracted the $0^\text{th}$-order position from a template fit, replacing the Fabry--Pérot template with the Argon \acl{SED} (\acs{SED}) template whenever the $0^\text{th}$-order was visible on the 2D spectrogram. This was primarily possible for acquisitions with the BGS000 grism. Contrary to this, the signal was below the detectable signal-to-noise ratio for the red grisms. Since we did not move the telescope simulator when changing the light source, we assumed for those cases that the $0^\text{th}$-order position of the Argon spectrogram was the same as that of the Fabry--Pérot spectrogram from the previous exposure.

\begin{figure}[!b]
        \centering
        \includegraphics[width=0.9\columnwidth]{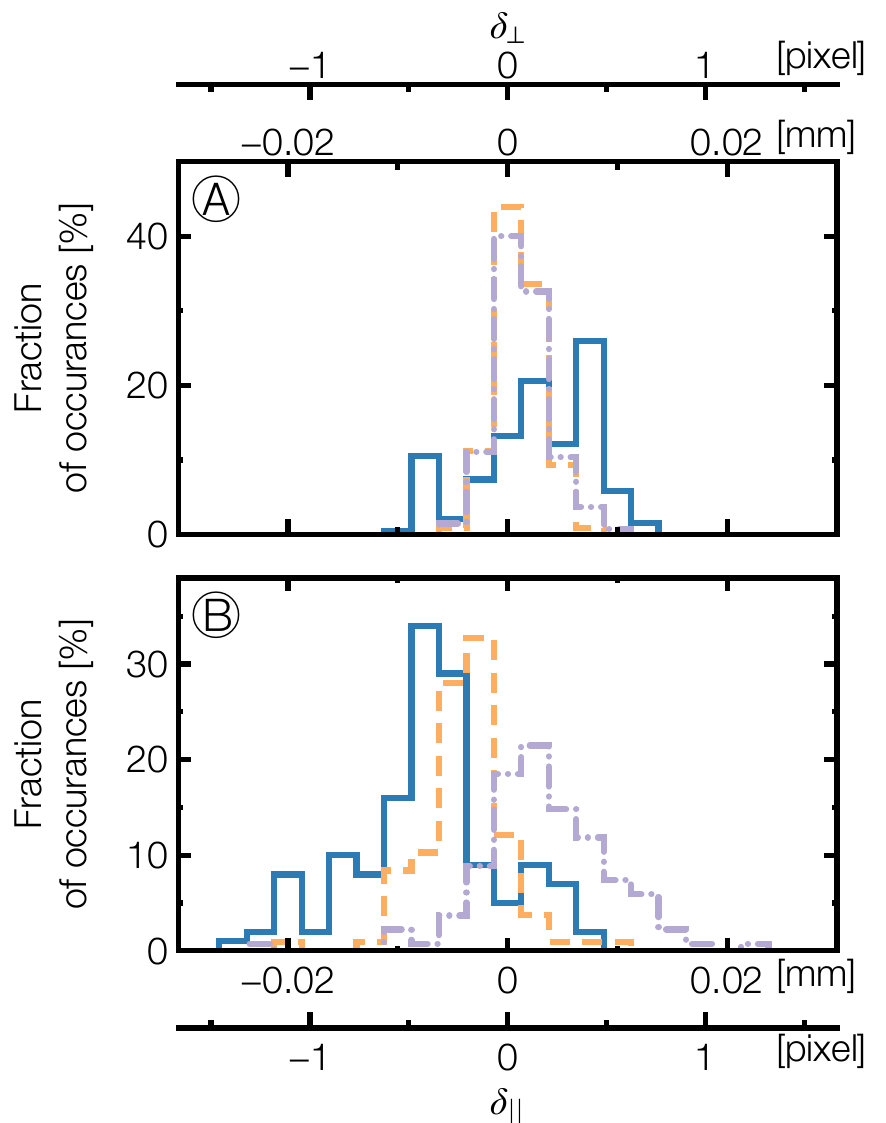}
        \caption{Histograms of the distances between measured and predicted Argon line positions across the \ac{NISP} field of view in the cross-dispersion direction $\delta_\perp$ (A) and in the dispersion direction $\delta_\parallel$ (B) -- for the Argon spectrograms captured with BGS000 (blue line) RGS000 (yellow dashed line) and RGS180 (purple dot-dashed line).  } \label{fig:Ar-check}
\end{figure} 

\begin{table*}[!t]
    \centering
       \caption{Average distance, standard deviation, and mean bias error between measured and predicted Argon line positions across NISP's field of view.}
       \begin{tabular}{c | c c c | c c c}
        \hline
             Grism   & \multicolumn{3}{c}{Cross-dispersion} & \multicolumn{3}{c}{Dispersion} \\
                   & Average [px] & Std-Dev [px] & Bias error [px] & Average [px] & Std-Dev [px] & Bias error [px]\\
                   \hline
               BGS000 &  $0.15\pm0.02$ & 0.29 & 0.33 & $-0.39\pm0.03$ & 0.36 & 0.53\\
               RGS000 &  $0.05\pm0.01$ & 0.12 & 0.13 & $-0.21\pm0.02$ & 0.21 & 0.29\\
               RGS180 &  $0.06\pm0.01$ & 0.14 & 0.15 & $ 0.19\pm0.03$ & 0.33 & 0.38\\
               \hline
        \end{tabular}
        \label{tab:Ar-check}   
\end{table*}

When inspecting Argon spectrograms captured with the BGS000, we found that four of them had very weak signal for the $0^\text{th}$-order resulting in the template fit failing to converge, positioning the $0^\text{th}$-order at about seven pixels away from visually determined position. These spectra were also excluded from our analyses.

Using calibration parameters that were derived from the Fabry--Pérot spectrogram analysis (see \cref{sc:FP-cal}), we estimated the position of the Argon emission lines relative to the $0^\text{th}$-order position and compared the predicted positions with the measured position. 

The bias resulting from this comparison is reported in \cref{tab:Ar-check}, defined as the average of the distribution as shown in \cref{fig:Ar-check}. It is induced by the persistence present in the images that tend to either bias the centroiding of the Argon lines themselves or the estimate of the position of the $0^\text{th}$-order.
As was described above, for the BGS000 an Argon \ac{SED} template was used to evaluate the $0^\text{th}$-order position, while for both red grisms this was taken from the previous exposure, assuming the telescope did not move between exposures. As a consequence, the weakness of the $0^\text{th}$-order and the persistence signal present in the BGS000 spectrogram leads to a larger bias than what is observed for the corresponding distribution of the red grisms.
Nonetheless, despite this bias reported in the \cref{tab:Ar-check}, we found that the predicted position of the Argon lines fall within a distance of $\simeq0.5$ pixels of the measured lines, which validates our calibration parametrisation.

\section{NISP Spectroscopic performance}
\label{sec:performance}

While calibration parameters have been computed on the basis of Chebyshev polynomials, it is more convenient to convert the obtained spectroscopic dispersion laws into regular polynomials to assess properties and performance of the grism's dispersion. In that case, the dispersion laws take the following form

\begin{equation}
        \begin{array}{r@{\,=\,}l}
            \displaystyle P_y = y-y_0 & \displaystyle \sum_{i=0}^{m-1} w^y_i(y_0,z_0)\,\lambda^i\,, \\
            \displaystyle P_z = z-z_0 & \displaystyle \sum_{j=0}^{n-1} w^z_i(y_0,z_0)\,\lambda^j\,,
        \end{array}
    \label{eq:regP_disperss}
\end{equation}
where $w^\kappa_i(y_0,z_0)$, with $\kappa$ being either $y$ or $z$, are the regular polynomial coefficients in cross-dispersion ($\kappa=y$) and dispersion direction ($\kappa=z$).
When converted to regular polynomials, the coefficients of zero degree ($w^\kappa_0$) are related to the Cartesian distance that separate the $1^\text{st}$-order of the spectra to the reference used in this work, i.e.\ the position of the $0^\text{th}$-order. 
The first degree coefficients ($w^\kappa_1$) are related to the dispersion $\mathrm{d}\lambda/\mathrm{d}\kappa$ of the spectra while coefficients of higher degree define the curvature of the spectra, which is linked to the non-linearity of the dispersion law and optical distortions of the system.

For clarity and simplicity, the forthcoming discussion will be confined to the description of the dispersion law averaged over the NISP \ac{FoV}. \cref{tab:to_regular} reports the dispersion coefficients in both the Chebyshev and regular basis and reveals that when the grisms are centred onto their nominal position, the cross-dispersion coefficient $w^y_1$ is about two orders of magnitude lower than the dispersion coefficient $w^z_1$. This aligns with the intended design of the grisms which primarily focuses on dispersing light in a single direction. Optical and mechanical distortions as well as the manufacturing limitations contribute to the distortion of the spectral track in both directions. However, even with those limitations, the higher-order coefficients ($w^\kappa_i$ with $i>1$) are small enough for the dispersion to be nearly a linear dispersion law mostly directed towards the $z$ axis.

In contrast, when the grisms are tilted by $\pm\ang{4}$, the cross-dispersion coefficient $w^y_1$ increases to be one order of magnitude lower than the dispersion coefficient $w^z_1$. This discrepancy with the nominal position arises due to the dispersion occurring in both directions when the dispersion grating is rotated by $\pm\ang{4}$.

Under the formalism used in this work, the dispersion of NISP's grisms predominantly takes place in the $z$ direction. As a consequence, our focus will now shift to the discussion of the regular $P_z$ polynomials.

In the context of slitless spectroscopy, spectral resolution is defined by a combination of both the intrinsic instrument \ac{PSF} as well as the characteristics of the targeted object's size, shape, and surface brightness distribution. Referring to the definition provided in \Euclid's scientific requirement, we define NISP's resolving power $\mathcal{R}$ as

\begin{equation}
    \mathcal{R} = \frac{\lambda\sqrt{2\ln{2}}}{\Delta \lambda\,\text{FWHM}(\lambda)}=\frac{\lambda}{2\,\Delta \lambda\,\sigma_e(\lambda)}\,.
    \label{eq:R}
\end{equation} 

In the above equation, $\Delta \lambda$ corresponds to the per-pixel spectral dispersion along the dispersion direction\footnote{along the $z$-axis for RGS000 and RGS180 in nominal position and in a mixture of $y$ and $z$ axis when those grisms are at $\pm4^\circ$ inclination} and is expressed in \micron/{px}. The $\text{FWHM}=2\sigma_e\sqrt{2\ln2}$ corresponds to the full-width-half-maximum of the target source with an axisymmetric Gaussian profile of effectives variance $\sigma_e^2$. This effective full-width-half-maximum results from the convolution between the luminosity profile of the source and the \ac{PSF} and can be expressed as:

\begin{equation}
    {\text{FWHM}}(\lambda)=\sqrt{\text{FWHM}^2_\mathrm{PSF}(\lambda)+\text{FWHM}^2_\mathrm{src}}\,.
    \label{eq:fwhm}
\end{equation} 
The $\text{FWHM}_ \mathrm{PSF}(\lambda)$ is the full-width-half-maximum of the \ac{PSF}, in pixels, at a given wavelength $\lambda$, and $\text{FWHM}_\mathrm{src}$ corresponds to the full-width-half-maximum, in pixels, of the target and depends on the source's surface brightness distribution.

When considering grisms RGS000 and RGS180 in their nominal position, the regular polynomial $P_z(\lambda)$ given in \cref{eq:regP_disperss} describes the parametrisation for the grism's dispersion law in the dispersion direction.
Its first-order coefficient $w_1^z$ can be related to the grism's dispersion $\mathrm{d} z/\mathrm{d}\lambda$.
To obtain an estimate of the per-pixel dispersion $\Delta\lambda=\mathrm{d} \lambda/\mathrm{d}z$ one needs to first invert this polynomial.
Given our parametrisation,  $P_z(\lambda)$ is continuously differentiable and non-null in the bandpass of the grisms, hence $P_z(\lambda)$ is invertible in the neighbourhood of $\lambda$, with its inverse being also differentiable at $b=P_z(\lambda)$ such that
\begin{equation}
    \left(P^{-1}_z\right)'(b)=\frac{1}{P_z'(\lambda)}\,.
\end{equation}

\begin{figure}[!t]
    \centering
        \begin{subfigure}{0.49\textwidth}
        \centering
                \caption{Blue grism BGS000 dispersion at 1\,\micron}\includegraphics[scale=0.6]{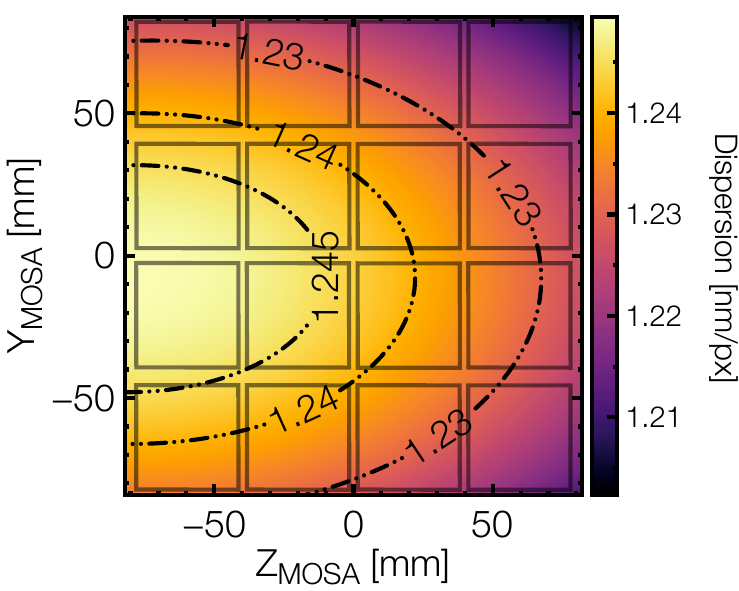} 
        \end{subfigure}
        \begin{subfigure}{0.49\textwidth}
        \centering
                \caption{Red grism RGS000 dispersion at 1.5\,\micron}\includegraphics[scale=0.6]{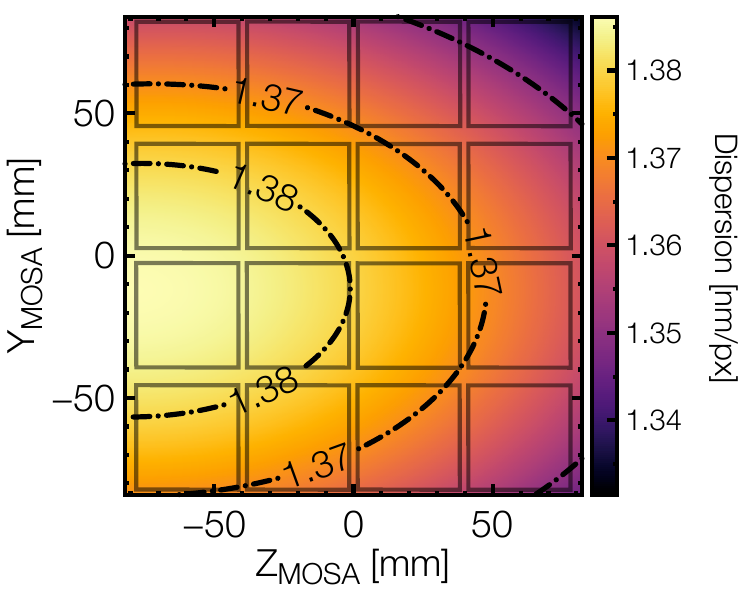}
        \end{subfigure}
        \begin{subfigure}{0.49\textwidth}
        \centering
                \caption{Red grism RGS180 dispersion at 1.5\,\micron}\includegraphics[scale=0.6]{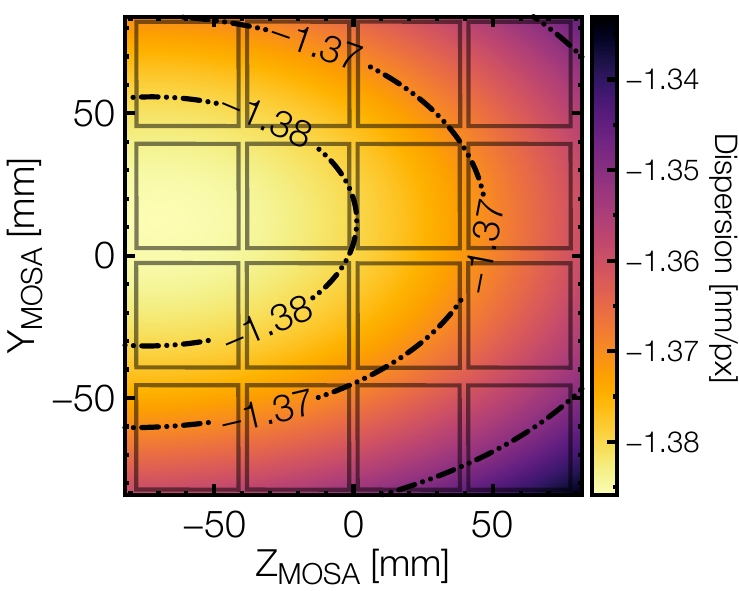}
        \end{subfigure}
        \caption{NISP's grism dispersion across the NISP focal plane reference frame identified as $\mathcal{R}_\text{MOSA}$ with MOSA standing for the pixel mosaic. In each figure, the squares represent the physical positions of the NISP detectors. In this representation we see the focal plane as an incoming photon will see it.\label{fig:2DDispersion}}
\end{figure}

The constant term in the derivative of the inverse function $(P^{-1}_z)'$ gives an estimate of the dispersion $\Delta \lambda$ at a wavelength $\lambda$ within in the grism's bandpass.  

\Cref{fig:2DDispersion} presents the variation across the field-of-view of the dispersion coefficients at the mid-bandpass wavelength, for all combinations of grism/rotation positions that were tested. The dispersion smoothly varies across the NISP \ac{FoV} following a radial gradient with its maximum, or minimum, on the left-hand side of the focal plane at about mosaic coordinates $(-80,\,0)\,\si{\milli\meter}$.
Notice the difference between the two red grisms: RGS000 has an average dispersion of the order of $+1.37\,\si{\nano\meter}\,\text{px}^{-1}$ while RGS180's average dispersion is of the order of $-1.37\,\si{\nano\meter}\,\text{px}^{-1}$.
The negative sign of the RGS180 dispersion originates from its design which imposes a dispersion of the light in the opposite direction from RGS000.
But on overall, RGS000 and RGS180 have a very similar absolute dispersion in terms of value and profile over the focal plane array, leading to a similar performance for both grisms. Note that, to highlight these similarities and due to the negative dispersion value of RGS180, colour scale for RGS180 is reversed relative to that of RGS000. In contrast, BGS000 has a lower dispersion with an average value of the order of $1.24\,\si{\nano\meter}\,\text{px}^{-1}$.
All these values are compatible with the design-specifications for each grism. 

\begin{figure}[!t]
    \centering
    \includegraphics[width=\columnwidth]{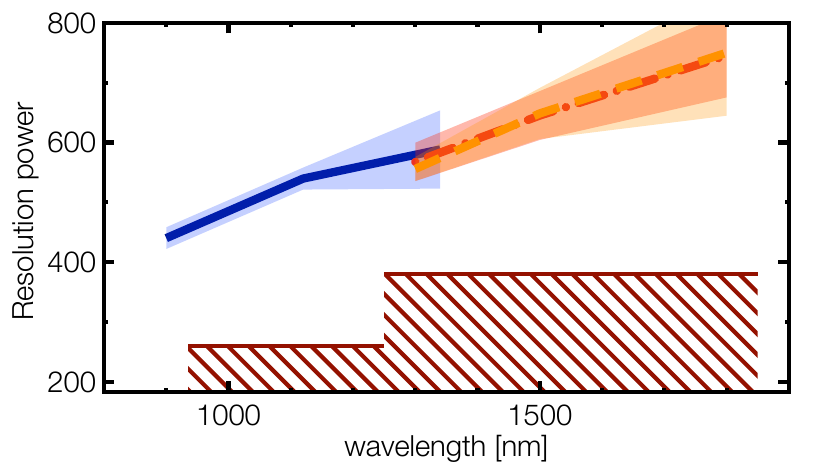}
    \caption{NISP's averaged resolving power for the BGS000 (line), RGS000 (dashed), and RGS180 (dot-dashed) grisms as function of wavelength, assuming a source size of 0\farcs5 \ac{FWHM}. The shaded areas illustrate the variation of the resolving power due to the variation of \ac{PSF} and the variation of the dispersion $\Delta \lambda$ across the \ac{FoV}. The hatched region shows the `forbidden' region below the scientific requirements.  } \label{fig:Resolving}
\end{figure}

Combining this estimate of the dispersion with the measured \ac{PSF} size, NISP's resolving power can be derived using \cref{eq:R,eq:fwhm}. The result is summarized in \cref{fig:Resolving} which shows NISP's averaged resolving power for a source of 0\farcs5 \ac{FWHM}.
The averaged resolving power from the blue grism increases from about $\mathcal{R} \simeq 440$ at $900\,\si{\nano\meter}$ to $\simeq690$ at $1300\,\si{\nano\meter}$, while both red grisms see their resolving power increase from $\mathcal{R}\simeq550$ at $1300\,\si{\nano\meter}$ to $\simeq740$ at $1800\,\si{\nano\meter}$.
These values have to be compared to the NISP scientific requirements of a resolving power $\mathcal{R} \geq 260$ for the blue grism and $\mathcal{R} \geq 380$ for the red grisms  and for a typical source size of $\ang{;;0.5}$. The combination of the high optical quality of the grisms and the NISP camera and collimator lenses ensures that NISP fulfils its requirements in terms of spectral dispersion and resolving power. 

In addition to the measurements described in this work, ground tests were conducted to verify compliance with stray light requirements from in-field and out-of-field sources. These confirmed that parasitic signals are below scientific performance thresholds. No further characterization was done, as the objective was limited to performance verification and time dedicated to the test campaign was constrained.

Again, shorter tests were conducted at the payload module level with the addition of the \Euclid mirrors and dichroic element to the light path. This allowed us to verify that the addition of the folding mirrors  as well as the transmission of the near-infrared light beam through the dichroic did not significantly modify NISP's spectral dispersion, resolving power as well as optical performances. 

\section{Conclusions and available data products}
\label{sec:conclusions}

The NISP instrument integration was completed in 2019. Before its delivery to \acs{ESA} for its integration into the payload module of the \Euclid spacecraft, the NISP instrument underwent a series of tests in a vacuum and cold environment to validate its operation mode as well as to validate its performance in view of the forthcoming flight operation. During the test campaigns that took place in the beginning of 2020 the optical performance of NISP were verified.

One of the measurement campaigns conducted under vacuum at cold temperatures was dedicated to the verification of the image quality provided by NISP's grisms as well as to the verification and validation of their dispersion. These are key properties in the accurate determination of cosmological redshifts which sits at the core of the derivation of cosmological parameters in galaxy clustering analyses.

While a non-conformity was revealed for the grism RGS270, which resulted in the modification of the \Euclid survey strategy, the analysis of the ground dataset presented in this work shows that NISP's grisms BGS000, RGS000, and RGS180 have an extremely high image quality with a \ac{PSF} encircled energy size well below the scientific requirements and very close to that of an ideal, solely diffraction-limited instrument. 

Secondly, these tests allowed us to verify that the spectral dispersion of NISP lies above requirements and showed that NISP's resolving power is larger than $\mathcal{R}= 480$ from \SI{1000}{\nano\meter} to \SI{1800}{\nano\meter}.
The RGS000, RGS180, RGS270, and BGS000 throughput for the $0^\text{th}$ and $1^\text{st}$ spectral orders at the centre of the NISP focal plane array are available at an \acs{ESA} server\footnote{\url{https://euclid.esac.esa.int/msp/refdata/nisp/NISP-SPECTRO-PASSBANDS-V1}}.
Available throughput numbers include contributions from the grisms, the telescope, and \ac{NI-OA}, as well as the mean detector quantum efficiency. For convenience, the tables are available in both ASCII and FITS format.

\begin{acknowledgements} 
\AckEC
\end{acknowledgements}

\bibliographystyle{aa}
\bibliography{NISP-S_perfo}

\begin{appendix}
\section{On ground calibration coefficients}
\label{app:calib}

\Cref{tab:to_regular} presents the averaged calibration coefficients used to model the spectral dispersion of the NISP grism. The table provides the Chebyshev polynomial coefficients directly derived from a fit of \cref{eq:2Dcal} to the ground dataset for all grisms but RGS270 and averaged over the NISP focal plane surface. For convenience, the table provides as well the equivalent averaged coefficients for regular polynomials. Those were obtained through the analytical transformation from the Chebyshev basis into the regular polynomial basis.

\Cref{tab:calib_BGS000_0} to \ref{tab:calib_RGS180_+4} give the $a_{ij}$ calibration coefficients for \cref{eq:2Dcal}, used to model spatial dependency of the $C^\kappa_i$ coefficient needed in \cref{eq:spectra}.

\begin{sidewaystable}
    \centering
    \def\arraystretch{1.25}
\caption{Averaged calibration coefficients given in the Chebyshev basis $ C^{\kappa}_i$ and in their regular polynomial basis $w^\kappa_i$}
\label{tab:to_regular}
\begin{tabular}{llrrrl|rrrl}
\hline
\hline
\multirow{2}{*}{Grism} & \multirow{2}{*}{Coef.} & \multicolumn{4}{c}{Chebyshev polynomial coefficients} & \multicolumn{4}{c}{Regular polynomial coefficient} \\
& & $\langle  C^{\kappa}_0\rangle_{yz}$ & $\langle  C^{\kappa}_1\rangle_{yz}$ & $\langle  C^{\kappa}_2\rangle_{yz}$ & $\langle  C^{\kappa}_3\rangle_{yz}$ & $\langle w^\kappa_0\rangle_{yz} [\text{mm}]$ & $\langle w^\kappa_1\rangle_{yz} [\text{mm}/\text{nm}]$ & $\langle w^\kappa_2\rangle_{yz} [\text{mm}/\text{nm}^2]$ & $\langle w^\kappa_3\rangle_{yz} [\text{mm}/\text{nm}^3]$ \\
\hline
\multirow{2}{*}{BGS000}    & $C^y$ & $6.7399\times10^{-2}$ & $1.5603\times10^{-2}$ & $-1.9453\times10^{-4}$ & & $6.7399\times10^{-2}$ & $1.5603\times10^{-2}$ & $-1.9453\times10^{-4}$ & \\
                           & $C^z$ & $1.5480\times10^{1}$ & $3.5725$ & $-2.0014\times10^{-3}$ & & $1.5480\times10^{1}$ & $3.5725$ & $-2.0014\times10^{-3}$ & \\
 \multirow{2}{*}{RGS000-4} & $C^y$ & $1.4183$ & $3.3267\times10^{-1}$ & $2.4864\times10^{-4}$ & & $1.4183$ & $3.3267\times10^{-1}$ & $2.4864\times10^{-4}$ & \\
                           & $C^z$ & $1.9622\times10^{1}$ & $4.5898$ & $4.6133\times10^{-3}$ & $8.5595\times10^{-4}$ & $1.9622\times10^{1}$ & $4.5898$ & $4.6133\times10^{-3}$ & $8.5595\times10^{-4}$ \\
\multirow{2}{*}{RGS000}    & $C^y$ & $5.0701\times10^{-2}$ & $1.2487\times10^{-2}$ & $-8.6940\times10^{-5}$ & & $5.0701\times10^{-2}$ & $1.2487\times10^{-2}$ & $-8.6940\times10^{-5}$ & \\
                           & $C^z$ & $1.9675\times10^{1}$ & $4.6058$ & $4.6519\times10^{-3}$ & $9.4801\times10^{-4}$ & $1.9675\times10^{1}$ & $4.6058$ & $4.6519\times10^{-3}$ & $9.4801\times10^{-4}$ \\
\multirow{2}{*}{RGS000+4}  & $C^y$ & $-1.2965$ & $-3.0384\times10^{-1}$ & $-3.5768\times10^{-4}$ & & $-1.2965$ & $-3.0384\times10^{-1}$ & $-3.5768\times10^{-4}$ & \\
                           & $C^z$ & $1.9636\times10^{1}$ & $4.6001$ & $4.4149\times10^{-3}$ & $8.5569\times10^{-4}$ & $1.9636\times10^{1}$ & $4.6001$ & $4.4149\times10^{-3}$ & $8.5569\times10^{-4}$ \\
\multirow{2}{*}{RGS180-4}  & $C^y$ & $-1.4395$ & $-3.3623\times10^{-1}$ & $2.0877\times10^{-4}$ & & $-1.4395$ & $-3.3623\times10^{-1}$ & $2.0877\times10^{-4}$ & \\
                           & $C^z$ & $-1.9634\times10^{1}$ & $-4.5974$ & $-4.2116\times10^{-3}$ & $-1.1804\times10^{-3}$ & $-1.9634\times10^{1}$ & $-4.5974$ & $-4.2116\times10^{-3}$ & $-1.1804\times10^{-3}$ \\
 \multirow{2}{*}{RGS180}   & $C^y$ & $-7.0567\times10^{-2}$ & $-1.5815\times10^{-2}$ & $3.4318\times10^{-4}$ & & $-7.0567\times10^{-2}$ & $-1.5815\times10^{-2}$ & $3.4318\times10^{-4}$ & \\
                           & $C^z$ & $-1.9679\times10^{1}$ & $-4.6050$ & $-4.2045\times10^{-3}$ & $-7.2379\times10^{-4}$ & $-1.9679\times10^{1}$ & $-4.6050$ & $-4.2045\times10^{-3}$ & $-7.2379\times10^{-4}$ \\
\multirow{2}{*}{RGS180+4}  & $C^y$ & $1.3016$ & $3.0505\times10^{-1}$ & $5.8173\times10^{-4}$ & & $1.3016$ & $3.0505\times10^{-1}$ & $5.8173\times10^{-4}$ & \\
                           & $C^z$ & $-1.9633\times10^{1}$ & $-4.5914$ & $-4.7848\times10^{-3}$ & $-1.0509\times10^{-3}$ & $-1.9633\times10^{1}$ & $-4.5914$ & $-4.7848\times10^{-3}$ & $-1.0509\times10^{-3}$ \\
\hline
\end{tabular}
\end{sidewaystable}

\begin{table*}
    \small
    \centering
    \def\arraystretch{1.5}
    \caption{Value of the BGS000 calibration parameters $a_{ij}$ that have to be used in \cref{eq:2Dcal} to model the large scale variation of the coefficient $C^\kappa_i$ of \cref{eq:spectra}. Cells with `$-$' indicate that we did not extend the Chebyshev expansion to that order level and corresponding $a_{ij}$ coefficients were not computed.}
    \label{tab:calib_BGS000_0}
    \begin{tabular}{ccccccc}
    \hline
    \hline
        Coef. &         $C^y_0$ &          $C^y_1$ &          $C^y_2$ &          $C^z_0$ &          $C^z_1$ &          $C^z_2$ \\
    \hline
    $a_{00}$ & $6.7399\times10^{-2}$ &  $1.5671\times10^{-2}$ & $-1.9453\times10^{-4}$ &  $1.5531\times10^{ 1}$ &  $3.5844\times10^{ 0}$ & $-2.0013\times10^{-3}$ \\
    $a_{01}$ & $1.5377\times10^{-2}$ &  $3.4106\times10^{-3}$ & $-9.2970\times10^{-5}$ & $-9.7090\times10^{-3}$ &  $1.1044\times10^{-2}$ &  $7.8402\times10^{-4}$ \\
    $a_{02}$ &           $-$&  $3.7860\times10^{-5}$ &            $-$&  $1.2165\times10^{-1}$ &  $2.8440\times10^{-2}$ &            $-$\\
    $a_{10}$ & $1.6054\times10^{-3}$ &  $6.4469\times10^{-3}$ & $-2.4419\times10^{-4}$ &  $1.4882\times10^{-1}$ &  $3.4848\times10^{-2}$ &            $-$\\
    $a_{11}$ & $1.5582\times10^{-1}$ &  $3.6446\times10^{-2}$ &  $4.3425\times10^{-4}$ & $-1.9311\times10^{-3}$ &  $2.1610\times10^{-5}$ &            $-$\\
    $a_{12}$ &           $-$& $-4.2604\times10^{-2}$ &            $-$& $-1.4697\times10^{-1}$ & $-3.5140\times10^{-2}$ &            $-$\\
    $a_{20}$ &           $-$&  $1.5113\times10^{-4}$ &            $-$&  $4.2428\times10^{-2}$ &  $9.5201\times10^{-3}$ &            $-$\\
    $a_{21}$ &           $-$& $-3.0894\times10^{-4}$ &            $-$&  $3.6337\times10^{-3}$ &  $5.2306\times10^{-4}$ &            $-$\\
    $a_{22}$ &           $-$&  $6.7420\times10^{-5}$ &            $-$& $-4.3685\times10^{-2}$ & $-9.7348\times10^{-3}$ &            $-$\\
    \hline
    \end{tabular}
\end{table*}

\begin{table*}
    \small
    \centering
    \def\arraystretch{1.5}
    \caption{Value of the RGS000--4$^\circ$ calibration parameters $a_{ij}$ that have to be used in \cref{eq:2Dcal} to model the large scale variation of the coefficient $C^\kappa_i$ of \cref{eq:spectra}. Cells with `$-$' indicate that we did not expend the Chebyshev expansion to that order level and corresponding $a_{ij}$ coefficients were not computed.}
    \label{tab:calib_RGS000_-4}
    \begin{tabular}{cccccccc}
    \hline
    \hline
        Coef. &         $C^y_0$ &         $C^y_1$ &          $C^y_2$ &          $C^z_0$ &          $C^z_1$ &         $C^z_2$ &         $C^z_3$ \\
    \hline
    $a_{00}$ & $1.4183\times10^{ 0}$ & $3.3267\times10^{-1}$ &  $2.4864\times10^{-4}$ &  $1.9687\times10^{ 1}$ &  $4.6052\times10^{ 0}$ & $4.6133\times10^{-3}$ & $8.5595\times10^{-4}$ \\
    $a_{01}$ & $1.4256\times10^{-2}$ & $4.8468\times10^{-3}$ &             $-$ &  $3.1117\times10^{-2}$ &  $2.4107\times10^{-2}$ & $1.4370\times10^{-3}$ & $2.6410\times10^{-5}$ \\
    $a_{02}$ &            $-$ &            $-$ &             $-$ &  $1.5528\times10^{-1}$ &  $3.6582\times10^{-2}$ &            $-$ &            $-$ \\
    $a_{03}$ &            $-$ &            $-$ &             $-$ & $-8.4918\times10^{-4}$ &             $-$ &            $-$ &            $-$ \\
    $a_{10}$ & $5.3844\times10^{-2}$ & $1.8815\times10^{-2}$ & $-3.8706\times10^{-4}$ &  $1.9144\times10^{-1}$ &  $4.5418\times10^{-2}$ & $2.3560\times10^{-5}$ & $9.6200\times10^{-6}$ \\
    $a_{11}$ & $1.3039\times10^{-1}$ & $2.7367\times10^{-2}$ &             $-$ &  $1.5371\times10^{-2}$ &  $3.8326\times10^{-3}$ & $5.9380\times10^{-5}$ & $7.0280\times10^{-5}$ \\
    $a_{12}$ &            $-$ &            $-$ &             $-$ &  $4.0145\times10^{-3}$ & $-4.9004\times10^{-2}$ &            $-$ &            $-$ \\
    $a_{13}$ &            $-$ &            $-$ &             $-$ & $-2.0956\times10^{-1}$ &             $-$ &            $-$ &            $-$ \\
    $a_{20}$ &            $-$ &            $-$ &             $-$ &  $5.0628\times10^{-2}$ &  $1.2502\times10^{-2}$ &            $-$ &            $-$ \\
    $a_{21}$ &            $-$ &            $-$ &             $-$ & $-2.2093\times10^{-3}$ &  $1.1722\times10^{-4}$ &            $-$ &            $-$ \\
    $a_{22}$ &            $-$ &            $-$ &             $-$ & $-1.4256\times10^{-3}$ & $-1.2620\times10^{-2}$ &            $-$ &            $-$ \\
    $a_{23}$ &            $-$ &            $-$ &             $-$ & $-5.0106\times10^{-2}$ &             $-$ &            $-$ &            $-$ \\
    $a_{30}$ &            $-$ &            $-$ &             $-$ &  $1.9669\times10^{-3}$ &             $-$ &            $-$ &            $-$ \\
    $a_{31}$ &            $-$ &            $-$ &             $-$ &  $6.0078\times10^{-4}$ &             $-$ &            $-$ &            $-$ \\
    $a_{32}$ &            $-$ &            $-$ &             $-$ &  $2.3606\times10^{-3}$ &             $-$ &            $-$ &            $-$ \\
    $a_{33}$ &            $-$ &            $-$ &             $-$ & $-4.0399\times10^{-3}$ &             $-$ &            $-$ &            $-$ \\
    \hline
    \end{tabular}
\end{table*}

\begin{table*}
    \small
    \centering
    \def\arraystretch{1.5}
    \caption{Value of the RGS000 calibration parameters $a_{ij}$ that have to be used in \cref{eq:2Dcal} to model the large scale variation of the coefficient $C^\kappa_i$ of \cref{eq:spectra}. Cells with `$-$' indicate that we did not expend the Chebyshev expansion to that order level and corresponding $a_{ij}$ coefficients were not computed.}
    \label{tab:calib_RGS000_0}
    \begin{tabular}{cccccccc}
    \hline
    \hline
        Coef &          $C^y_0$  &          $C^y_1$  &           $C^y_2$  &           $C^z_0$  &           $C^z_1$  &          $C^z_2$  &           $C^z_3$  \\
    \hline
    $a_{00}$ & $5.0701\times10^{-2}$ & $1.2487\times10^{-2}$ & $-8.6940\times10^{-5}$ &  $1.9740\times10^{1}$ &  $4.6211\times10^{0}$ & $4.6519\times10^{-3}$ &  $9.4801\times10^{-4}$ \\
    $a_{01}$ & $2.0709\times10^{-2}$ & $4.7953\times10^{-3}$ &             $-$ &  $2.1814\times10^{-2}$ &  $2.1329\times10^{-2}$ & $1.1273\times10^{-3}$ & $-1.4514\times10^{-4}$ \\
    $a_{02}$ &            $-$ &            $-$ &             $-$ &  $1.5542\times10^{-1}$ &  $3.6585\times10^{-2}$ &            $-$ &             $-$ \\
    $a_{03}$ &            $-$ &            $-$ &             $-$ & $-3.8228\times10^{-4}$ &             $-$ &            $-$ &             $-$ \\
    $a_{10}$ & $3.3689\times10^{-2}$ & $1.3301\times10^{-2}$ & $-3.9825\times10^{-4}$ &  $1.9353\times10^{-1}$ &  $4.4968\times10^{-2}$ &            $-$ &             $-$ \\
    $a_{11}$ & $1.6763\times10^{-1}$ & $3.3932\times10^{-2}$ &             $-$ &  $3.2088\times10^{-3}$ &  $2.7488\times10^{-4}$ &            $-$ &             $-$ \\
    $a_{12}$ &            $-$ &            $-$ &             $-$ &  $3.3065\times10^{-3}$ & $-4.5248\times10^{-2}$ &            $-$ &             $-$ \\
    $a_{13}$ &            $-$ &            $-$ &             $-$ & $-1.9875\times10^{-1}$ &             $-$ &            $-$ &             $-$ \\
    $a_{20}$ &            $-$ &            $-$ &             $-$ &  $5.1402\times10^{-2}$ &  $1.2537\times10^{-2}$ &            $-$ &             $-$ \\
    $a_{21}$ &            $-$ &            $-$ &             $-$ &  $1.1812\times10^{-3}$ &  $9.6595\times10^{-4}$ &            $-$ &             $-$ \\
    $a_{22}$ &            $-$ &            $-$ &             $-$ & $-2.6818\times10^{-3}$ & $-1.3090\times10^{-2}$ &            $-$ &             $-$ \\
    $a_{23}$ &            $-$ &            $-$ &             $-$ & $-5.3760\times10^{-2}$ &             $-$ &            $-$ &             $-$ \\
    $a_{30}$ &            $-$ &            $-$ &             $-$ &  $1.6468\times10^{-3}$ &             $-$ &            $-$ &             $-$ \\
    $a_{31}$ &            $-$ &            $-$ &             $-$ &  $2.6921\times10^{-3}$ &             $-$ &            $-$ &             $-$ \\
    $a_{32}$ &            $-$ &            $-$ &             $-$ &  $1.4953\times10^{-3}$ &             $-$ &            $-$ &             $-$ \\
    $a_{33}$ &            $-$ &            $-$ &             $-$ & $-5.6130\times10^{-3}$ &             $-$ &            $-$ &             $-$ \\
    \hline
    \end{tabular}
\end{table*}

\begin{table*}
    \small
    \centering
    \def\arraystretch{1.5}
    \caption{Value of the RGS000+4$^\circ$ calibration parameters $a_{ij}$ that have to be used in \cref{eq:2Dcal} to model the large scale variation of the coefficient $C^\kappa_i$ of \cref{eq:spectra}. Cells with `$-$' indicate that we did not expend the Chebyshev expansion to that order level and corresponding $a_{ij}$ coefficients were not computed.}
    \label{tab:calib_RGS000_+4}
    \begin{tabular}{cccccccc}
    \hline
    \hline
        Coef. &          $C^y_0$ &          $C^y_1$ &          $C^y_2$ &          $C^z_0$ &          $C^z_1$ &          $C^z_2$ &          $C^z_3$ \\
    \hline
    $a_{00}$ & $-1.2965\times10^{ 0}$ & $-3.0475\times10^{-1}$ & $-3.5768\times10^{-4}$ &  $1.9701\times10^{ 1}$ &  $4.6157\times10^{ 0}$ &  $4.4149\times10^{-3}$ &  $8.5569\times10^{-4}$ \\
    $a_{01}$ &  $2.1410\times10^{-2}$ &  $3.9783\times10^{-3}$ &             $-$ &  $9.3928\times10^{-3}$ &  $1.9468\times10^{-2}$ &  $1.4553\times10^{-3}$ &  $1.0114\times10^{-4}$ \\
    $a_{02}$ &             $-$ & $-7.0219\times10^{-4}$ &             $-$ &  $1.5623\times10^{-1}$ &  $3.7125\times10^{-2}$ &             $-$ &             $-$ \\
    $a_{03}$ &             $-$ &             $-$ &             $-$ & $-1.7588\times10^{-3}$ &             $-$ &             $-$ &             $-$ \\
    $a_{10}$ &  $4.2449\times10^{-3}$ &  $7.7817\times10^{-3}$ & $-3.4587\times10^{-4}$ &  $1.9675\times10^{-1}$ &  $4.4186\times10^{-2}$ & $-1.1720\times10^{-4}$ & $-1.4462\times10^{-4}$ \\
    $a_{11}$ &  $2.1318\times10^{-1}$ &  $4.8050\times10^{-2}$ &             $-$ & $-2.9817\times10^{-3}$ & $-2.7084\times10^{-3}$ & $-5.5900\times10^{-5}$ &  $2.4555\times10^{-4}$ \\
    $a_{12}$ &             $-$ & $-5.5537\times10^{-2}$ &             $-$ &  $8.2906\times10^{-3}$ & $-4.1701\times10^{-2}$ &             $-$ &             $-$ \\
    $a_{13}$ &             $-$ &             $-$ &             $-$ & $-1.9329\times10^{-1}$ &             $-$ &             $-$ &             $-$ \\
    $a_{20}$ &             $-$ & $-2.2372\times10^{-3}$ &             $-$ &  $5.2079\times10^{-2}$ &  $1.2488\times10^{-2}$ &             $-$ &             $-$ \\
    $a_{21}$ &             $-$ & $-6.4550\times10^{-5}$ &             $-$ & $-5.3066\times10^{-3}$ &  $7.4314\times10^{-4}$ &             $-$ &             $-$ \\
    $a_{22}$ &             $-$ &  $2.1837\times10^{-3}$ &             $-$ & $-8.6373\times10^{-4}$ & $-1.2726\times10^{-2}$ &             $-$ &             $-$ \\
    $a_{23}$ &             $-$ &             $-$ &             $-$ & $-4.8376\times10^{-2}$ &             $-$ &             $-$ &             $-$ \\
    $a_{30}$ &             $-$ &             $-$ &             $-$ &  $3.8456\times10^{-3}$ &             $-$ &             $-$ &             $-$ \\
    $a_{31}$ &             $-$ &             $-$ &             $-$ &  $8.5105\times10^{-3}$ &             $-$ &             $-$ &             $-$ \\
    $a_{32}$ &             $-$ &             $-$ &             $-$ &  $4.8891\times10^{-3}$ &             $-$ &             $-$ &             $-$ \\
    $a_{33}$ &             $-$ &             $-$ &             $-$ & $-9.3452\times10^{-3}$ &             $-$ &             $-$ &             $-$ \\
    \hline
    \end{tabular}
\end{table*}

\begin{table*}
    \small
    \centering
    \def\arraystretch{1.5}
    \caption{Value of the RGS180--4 calibration parameters $a_{ij}$ that have to be used in \cref{eq:2Dcal} to model the large scale variation of the coefficient $C^\kappa_i$ of \cref{eq:spectra}. Cells with `$-$' indicate that we did not expend the Chebyshev expansion to that order level and corresponding $a_{ij}$ coefficients were not computed.}
    \label{tab:calib_RGS180_-4}
    \begin{tabular}{cccccccc}
    \hline
    \hline
        Coef &          $C^y_0$ &          $C^y_1$ &          $C^y_2$ &          $C^z_0$ &          $C^z_1$ &          $C^z_2$ &          $C^z_3$ \\
    \hline
    $a_{00}$ & $-1.4442\times10^{0}$ & $-3.3737\times10^{-1}$ &  $2.0877\times10^{-4}$ & $-1.9701\times10^{1}$ & $-4.6132\times10^{0}$ & $-4.2116\times10^{-3}$ & $-1.1804\times10^{-3}$ \\
    $a_{01}$ & $-2.0628\times10^{-2}$ & $-3.7792\times10^{-3}$ &             $-$ &  $1.1624\times10^{-2}$ &  $1.8410\times10^{-2}$ &  $1.1478\times10^{-3}$ & $-1.5622\times10^{-4}$ \\
    $a_{02}$ & $-5.3253\times10^{-3}$ & $-1.1466\times10^{-3}$ &             $-$ & $-1.6132\times10^{-1}$ & $-3.7370\times10^{-2}$ &             $-$ &             $-$ \\
    $a_{03}$ &             $-$ &             $-$ &             $-$ &  $5.5879\times10^{-3}$ &             $-$ &             $-$ &             $-$ \\
    $a_{10}$ &  $5.8618\times10^{-3}$ &  $7.2148\times10^{-3}$ & $-1.3234\times10^{-4}$ & $-1.9315\times10^{-1}$ & $-4.4419\times10^{-2}$ &  $2.1423\times10^{-4}$ & $-5.3528\times10^{-4}$ \\
    $a_{11}$ & $-2.0715\times10^{-1}$ & $-4.7939\times10^{-2}$ &             $-$ & $-1.2133\times10^{-2}$ & $-2.5801\times10^{-3}$ & $-1.8255\times10^{-4}$ &  $6.2993\times10^{-4}$ \\
    $a_{12}$ &  $1.9992\times10^{-1}$ &  $4.0851\times10^{-2}$ &             $-$ & $-4.4835\times10^{-3}$ &  $4.6811\times10^{-2}$ &             $-$ &             $-$ \\
    $a_{13}$ &             $-$ &             $-$ &             $-$ &  $2.0969\times10^{-1}$ &             $-$ &             $-$ &             $-$ \\
    $a_{20}$ & $-1.0219\times10^{-2}$ & $-2.3980\times10^{-3}$ &             $-$ & $-5.0410\times10^{-2}$ & $-1.3488\times10^{-2}$ &             $-$ &             $-$ \\
    $a_{21}$ &  $2.4316\times10^{-3}$ &  $3.9580\times10^{-5}$ &             $-$ &  $5.5960\times10^{-5}$ &  $7.4457\times10^{-4}$ &             $-$ &             $-$ \\
    $a_{22}$ &  $6.5568\times10^{-3}$ &  $2.5569\times10^{-3}$ &             $-$ & $-1.2132\times10^{-3}$ &  $1.1499\times10^{-2}$ &             $-$ &             $-$ \\
    $a_{23}$ &             $-$ &             $-$ &             $-$ &  $5.7432\times10^{-2}$ &             $-$ &             $-$ &             $-$ \\
    $a_{30}$ &             $-$ &             $-$ &             $-$ & $-9.1911\times10^{-4}$ &             $-$ &             $-$ &             $-$ \\
    $a_{31}$ &             $-$ &             $-$ &             $-$ &  $2.0260\times10^{-3}$ &             $-$ &             $-$ &             $-$ \\
    $a_{32}$ &             $-$ &             $-$ &             $-$ & $-4.4455\times10^{-3}$ &             $-$ &             $-$ &             $-$ \\
    $a_{33}$ &             $-$ &             $-$ &             $-$ &  $1.1781\times10^{-3}$ &             $-$ &             $-$ &             $-$ \\
    \hline
    \end{tabular}
\end{table*}

\begin{table*}
    \small
    \centering
    \def\arraystretch{1.5}
    \caption{Value of the RGS180 calibration parameters $a_{ij}$ that have to be used in \cref{eq:2Dcal} to model the large scale variation of the coefficient $C^\kappa_i$ of \cref{eq:spectra}. Cells with `$-$' indicate that we did not expend the Chebyshev expansion to that order level and corresponding $a_{ij}$ coefficients were not computed.}
    \label{tab:calib_RGS180_0}
    \begin{tabular}{cccccccc}
    \hline
    \hline
        Coef &          $C^y_0$ &          $C^y_1$ &          $C^y_2$ &          $C^z_0$ &          $C^z_1$ &          $C^z_2$ &          $C^z_3$ \\
    \hline
    $a_{00}$ & $-7.0567\times10^{-2}$ & $-1.5815\times10^{-2}$ &  $3.4318\times10^{-4}$ & $-1.9747\times10^{1}$ & $-4.6205\times10^{0}$ & $-4.2045\times10^{-3}$ & $-7.2379\times10^{-4}$ \\
    $a_{01}$ & $-2.0116\times10^{-2}$ & $-4.4326\times10^{-3}$ &             $-$ &  $2.1571\times10^{-2}$ &  $2.0309\times10^{-2}$ &  $1.2335\times10^{-3}$ & $-1.1032\times10^{-4}$ \\
    $a_{02}$ &             $-$ &             $-$ &             $-$ & $-1.6139\times10^{-1}$ & $-3.6443\times10^{-2}$ &             $-$ &             $-$ \\
    $a_{03}$ &             $-$ &             $-$ &             $-$ &  $2.2445\times10^{-3}$ &             $-$ &             $-$ &             $-$ \\
    $a_{10}$ &  $3.0841\times10^{-2}$ &  $1.3088\times10^{-2}$ & $-2.5114\times10^{-4}$ & $-1.9029\times10^{-1}$ & $-4.4870\times10^{-2}$ &             $-$ &             $-$ \\
    $a_{11}$ & $-2.3678\times10^{-1}$ & $-6.0037\times10^{-2}$ &             $-$ &  $2.9505\times10^{-3}$ & $-6.3743\times10^{-4}$ &             $-$ &             $-$ \\
    $a_{12}$ &             $-$ &             $-$ &             $-$ & $-2.8491\times10^{-3}$ &  $4.5974\times10^{-2}$ &             $-$ &             $-$ \\
    $a_{13}$ &             $-$ &             $-$ &             $-$ &  $1.8862\times10^{-1}$ &             $-$ &             $-$ &             $-$ \\
    $a_{20}$ &             $-$ &             $-$ &             $-$ & $-5.6851\times10^{-2}$ & $-1.2420\times10^{-2}$ &             $-$ &             $-$ \\
    $a_{21}$ &             $-$ &             $-$ &             $-$ &  $3.4777\times10^{-3}$ & $-1.0028\times10^{-3}$ &             $-$ &             $-$ \\
    $a_{22}$ &             $-$ &             $-$ &             $-$ & $-2.4494\times10^{-3}$ &  $1.4623\times10^{-2}$ &             $-$ &             $-$ \\
    $a_{23}$ &             $-$ &             $-$ &             $-$ &  $6.1417\times10^{-2}$ &             $-$ &             $-$ &             $-$ \\
    $a_{30}$ &             $-$ &             $-$ &             $-$ &  $3.0686\times10^{-3}$ &             $-$ &             $-$ &             $-$ \\
    $a_{31}$ &             $-$ &             $-$ &             $-$ &  $1.4954\times10^{-3}$ &             $-$ &             $-$ &             $-$ \\
    $a_{32}$ &             $-$ &             $-$ &             $-$ &  $4.6541\times10^{-4}$ &             $-$ &             $-$ &             $-$ \\
    $a_{33}$ &             $-$ &             $-$ &             $-$ & $-8.8378\times10^{-3}$ &             $-$ &             $-$ &             $-$ \\
    \hline
    \end{tabular}
\end{table*}
    
\begin{table*}
    \small
    \centering
    \def\arraystretch{1.5}
    \caption{Value of the RGS180+4 calibration parameters $a_{ij}$ that have to be used in \cref{eq:2Dcal} to model the large scale variation of the coefficient $C^\kappa_i$ of \cref{eq:spectra}. Cells with `$-$' indicate that we did not expend the Chebyshev expansion to that order level and corresponding $a_{ij}$ coefficients were not computed.}
    \label{tab:calib_RGS180_+4}
    \begin{tabular}{cccccccc}
    \hline
    \hline
        Coef &          $C^y_0$ &          $C^y_1$ &          $C^y_2$ &          $C^z_0$ &          $C^z_1$ &          $C^z_2$ &          $C^z_3$ \\
    \hline
    $a_{00}$ &  $1.3061\times10^{0}$ &  $3.0601\times10^{-1}$ &  $5.8173\times10^{-4}$ & $-1.9699\times10^{1}$ & $-4.6069\times10^{0}$ & $-4.7848\times10^{-3}$ & $-1.0509\times10^{-3}$ \\
    $a_{01}$ & $-1.7548\times10^{-2}$ & $-4.8886\times10^{-3}$ &             $-$ &  $3.1209\times10^{-2}$ &  $2.3059\times10^{-2}$ &  $1.2071\times10^{-3}$ & $-1.8973\times10^{-4}$ \\
    $a_{02}$ &  $3.1466\times10^{-3}$ &  $7.7696\times10^{-4}$ &             $-$ & $-1.5818\times10^{-1}$ & $-3.6702\times10^{-2}$ &             $-$ &             $-$ \\
    $a_{03}$ &             $-$ &             $-$ &             $-$ &  $4.0210\times10^{-5}$ &             $-$ &             $-$ &             $-$ \\
    $a_{10}$ &  $5.7736\times10^{-2}$ &  $1.8916\times10^{-2}$ & $-3.8133\times10^{-4}$ & $-1.8310\times10^{-1}$ & $-4.4734\times10^{-2}$ & $-2.0143\times10^{-4}$ &  $1.4270\times10^{-5}$ \\
    $a_{11}$ & $-2.0123\times10^{-1}$ & $-4.6850\times10^{-2}$ &             $-$ &  $5.2091\times10^{-3}$ &  $3.4692\times10^{-3}$ &  $8.8720\times10^{-5}$ &  $1.4605\times10^{-4}$ \\
    $a_{12}$ &  $1.4235\times10^{-1}$ &  $2.7686\times10^{-2}$ &             $-$ &  $5.9618\times10^{-3}$ &  $4.1991\times10^{-2}$ &             $-$ &             $-$ \\
    $a_{13}$ &             $-$ &             $-$ &             $-$ &  $1.6494\times10^{-1}$ &             $-$ &             $-$ &             $-$ \\
    $a_{20}$ &  $1.1131\times10^{-2}$ &  $2.3656\times10^{-3}$ &             $-$ & $-5.0883\times10^{-2}$ & $-1.2707\times10^{-2}$ &             $-$ &             $-$ \\
    $a_{21}$ &  $8.7586\times10^{-4}$ & $-3.3949\times10^{-4}$ &             $-$ & $-4.9381\times10^{-3}$ & $-5.5980\times10^{-5}$ &             $-$ &             $-$ \\
    $a_{22}$ & $-1.2028\times10^{-2}$ & $-1.7466\times10^{-3}$ &             $-$ &  $2.6308\times10^{-3}$ &  $1.2065\times10^{-2}$ &             $-$ &             $-$ \\
    $a_{23}$ &             $-$ &             $-$ &             $-$ &  $5.1416\times10^{-2}$ &             $-$ &             $-$ &             $-$ \\
    $a_{30}$ &             $-$ &             $-$ &             $-$ &  $6.0270\times10^{-3}$ &             $-$ &             $-$ &             $-$ \\
    $a_{31}$ &             $-$ &             $-$ &             $-$ & $-1.1937\times10^{-2}$ &             $-$ &             $-$ &             $-$ \\
    $a_{32}$ &             $-$ &             $-$ &             $-$ &  $5.8415\times10^{-3}$ &             $-$ &             $-$ &             $-$ \\
    $a_{33}$ &             $-$ &             $-$ &             $-$ & $-6.8113\times10^{-3}$ &             $-$ &             $-$ &             $-$ \\
    \hline
    \end{tabular}
\end{table*}

\end{appendix}

\end{document}